\documentclass[11pt]{article} 
\pdfoutput=1 

\usepackage{jheppub} 

\usepackage[utf8]{inputenc}
\usepackage[english]{babel}
\usepackage{amsmath,amssymb,amsbsy,amstext,amsthm,booktabs,simplewick,exscale,relsize,slashed,graphicx,amsfonts,upgreek, xcolor,subcaption}
\usepackage{multirow,color,dcolumn,bm,enumerate}
\usepackage{hyperref}
\usepackage{feynmp-auto,axodraw2,tikz}
\usepackage{ulem}
\usepackage{comment}
\usepackage{cancel}

\usepackage{hyperref}
\usepackage[capitalise,noabbrev,nameinlink]{cleveref}

\title{Geometric Amplitudes: A Covariant Functional Approach for Massless Scalar Theories}
\author[a]{Antonio Delgado,}
\author[a]{Adam Martin,}
\author[b]{Runqing Wang}

\affiliation[a]{Department of Physics and Astronomy, University of Notre Dame,
  South Bend, IN, 46556 USA}
\affiliation[b]{Department of Physics and Center for Field Theory and Particle Physics,\\ Fudan University, Shanghai 200438, China}
\emailAdd{adelgad2@nd.edu}
\emailAdd{amarti41@nd.edu}
\emailAdd{runqing$\underline{~}$wang@fudan.edu.cn}

\abstract{
Functional geometry is a framework using concepts from geometry to understand the invariance of amplitudes in quantum field theory under a large class of field redefinitions, including those involving derivatives. It is inspired by recursion relations among correlation functions, where higher-point functions depend iteratively upon smaller correlators. Previous work has shown that, with suitable modifications, these correlation functions become covariant under field redefinitions, provided they are evaluated at the physical ``on-shell" point. In this paper, we show how to further modify correlation functions in massless scalar field theories to achieve ``off-shell" covariance. We investigate the conditions required for the framework to work and discuss the geometric interpretation of this construction -- which prioritizes the covariant transformation of observables under field redefinitions over the role of a metric tensor and its derivatives. While analogous modifications may exist for massive theories, we show that framework developed here does not extend straightforwardly to that case.}

\begin{document}
\maketitle
\setcounter{page}{2}

\section{Introduction}\label{intro} 

Effective field theories are used in different areas of physics to describe a variety of problems. One writes a Lagrangian (or Hamiltonian) with the relevant degrees of freedom at the energies of the system being studied. A typical formulation is to expand the Lagrangian in a series of operators, which comprise possible (and usually independent) interactions compatible with all symmetries of the theory. However, it has been known for a long time that this approach involves non-physical redundancies due to its off-shell nature. For example, field redefinitions \cite{Arzt:1993gz,Criado:2018sdb,Cohen:2024fak} and integration by parts allow for different parameterizations of the fields and interactions without affecting the physical scattering amplitudes.

These kinds of redundancy are not new in physics. For instance, it is believed that the laws of physics should be coordinate-independent, i.e. covariant under coordinate transformations. This realization is mostly achieved using differential geometry, which provides a natural description of physical observables in a basis-independent manner. This framework first appeared in the context of a pseudo-Riemannian manifold in the study of General Relativity, where two revolutionary ideas were synthesized: general covariance, which mandates coordinate-independent laws, and the existence of a dynamic metric, whose curvature manifests as the gravitational interaction. A similar story occurs in gauge theories, where different gauge choices lead to different effective actions, although they share the same on-shell physics. In the 1980s, Vilkovisky and DeWitt \cite{Vilkovisky:1984st,DeWitt1985, DeWitt1992Supermanifolds} constructed a unique geometric effective action whose goal was to eliminate all off-shell nonphysical ambiguities. The crucial point in the formalism is to identify a connection and the corresponding covariant derivative such that all quantities can be identified as  tensors, evaluated assuming on-shell conditions.  

The same insights exist in the study of geometric EFT amplitudes, where fields are identified with coordinates, and field redefinitions are treated as coordinate transformations acting on the underlying manifold. Amplitudes are related to covariant quantities and thus are manifestly field-basis independent. For example, let us consider the general two-derivative theory of scalar fields defined by 
\begin{equation}\label{eq: NLSM Lagrangian}
\mathcal L=\frac{1}{2}g_{ab}(\phi)\partial\phi^a\partial\phi^b,
\end{equation}
where $a,b$ are flavor indices. The positive-definite\footnote{As a result of unitarity.}and symmetric kinetic factor $g_{ab}(\phi)=g_{ba}(\phi)$ is a polynomial in $\phi$. Provided we restrict to field redefinitions that don't involve derivatives $\phi^i\rightarrow\phi'^i=\phi'^i(\phi)$, $g_{ab}$ transform in the following way
\begin{equation}\label{eq: g transform}
    g_{ab}(\phi)\rightarrow g'_{ab}(\phi')=g_{cd}(\phi)\frac{\partial\phi^c}{\partial\phi'^a}\frac{\partial\phi^d}{\partial\phi'^b}.
\end{equation}
Consequently, $g_{ab}$ can be recognized as a metric defined on  field-space. The 4-point amplitude\footnote{All 3-point amplitudes in the theory \eqref{eq: NLSM Lagrangian} vanish because they are derivatively interacted.} $\mathcal A_{abcd}$ can thus be related to the Riemann curvature $R_{abcd}$ naturally induced by the metric $g_{ab}(\phi)$, i.e. $\mathcal A_{abcd}=R_{abcd}u+R_{acbd}s$, where $u,s$ are Mandelstam variables in the $2\rightarrow2$ process. Since the curvature tensor is covariant by construction, these 4-point amplitudes are manifestly invariant under non-derivative field redefinitions. The underlying structure is known as  field-space geometry \cite{Alonso:2015fsp, Alonso:2016oah} and natural generalizations involving higher-spin fields have also been proposed by embedding field-space in a supermanifold for fermions~\cite{Finn:2020nvn, Assi:2023zid,Assi:2025fsm}, introducing two different metrics for gauge bosons~\cite{Helset:2022tlf} and adding Grassmann variables for supersymmetric extensions~\cite{DeWitt1992Supermanifolds,Finn:2020nvn, Gattus:2023gep, Gattus:2024ird,Lee:2024xqa}. Phenomenological studies of the geometric interpretation have been applied both to SMEFT and HEFT\cite{Cohen:2020xca, Cohen:2021ucp}, including the derivation of soft theorems \cite{Derda:2024jvo,Cohen:2025dex}, renormalizing field space geometry \cite{Aigner:2025xyt}, geometric matching \cite{Li:2024ciy}, and the renormalization group equations (RGE) applied to the Higgs sector~\cite{Helset:2022pde}.

Recently, there has been much interest in extending the geometric picture/interpretation to include field redefinitions that involve derivatives $\phi^i\rightarrow\phi'^i=\phi'^i(\phi,\partial_\mu\phi,\partial_\mu\partial_\nu\phi,\cdots)$. This is not a small adjustment, as it requires combining field space (where the fields are the coordinates of the manifold) with configuration space (spacetime coordinates $x^\mu$). If we wish to include redefinitions involving all derivative powers, this implies an infinite base manifold. There have been several different approaches in the literature towards making this leap. One approach is the jet bundle method~\cite{Alminawi:2023qtf,Alminawi:2025pwg, Craig:2023hhp, Craig:2025uoc} (or, slightly simpler but less general, a Lagrange space~\cite{Craig:2023wni}), where field derivatives are treated as new coordinates, and d-tensor derivatives preserve covariance by introducing non-linear connections when higher derivative powers are involved. A second approach is known as geometry-kinematics duality~\cite{Cheung:2022vnd}, which maps a generic scalar EFT onto NLSM (Eq. \eqref{eq: NLSM Lagrangian}) by working in momentum space. Amplitudes are related to the covariant kinetic metric and curvature, and thus becoming manifestly invariant under field redefinitions. The final approach is known as functional geometry, which parametrizes all possible configurations of a field defined over a given spacetime manifold, spanned by $\{\phi,\partial_\mu\phi,\partial_\mu\partial_\nu\phi,\cdots\}$. Partial derivatives ${\partial}/{\partial\phi^i}$ are replaced with functional derivatives ${\delta/}{\delta\phi^i}$ to maintain the vectorial property. The building blocks in this setup are correlation functions, which can be recursively related to each other via a connection formed from the two- and three-point functions. While current versions of the functional approach have a natural geometric interpretation on-shell, the interpretation breaks down off-shell. All of the above approaches share the starting point that physical quantities should be related to tensors under field redefinitions to  manifestly impose the invariance of scattering amplitudes. 

In this paper, we take the last approach, i.e. functional geometry, as the framework to build upon. Recent studies \cite{Cohen:2022uuw, Cohen:2023ekv} work with on-shell covariant building blocks, i.e. quantities at a certain point in the field configuration space, since they are directly related to physical observables. It has been shown that in a scalar EFT, the $n$-point correlation functions $\mathcal{M}_{12\cdots n}$ are on-shell covariant\footnote{A rigorous definition is given in Eq. \eqref{eq:onshellcov} and \eqref{eq:onshellcov2}.} under field redefinitions, and they can be recursively constructed from lower-point functions; while the off-shell covariance is spoiled by the existence of non-zero ``anholonomic" terms, which only vanish on-shell. 

Our aim is to re-examine the formulation, trying to find a covariant off-shell recursion relation that relates higher-point amplitudes to lower-point amplitudes. In our approach, even though obtaining an on-shell result is the ultimate goal, we will take intermediate off-shell steps that can be useful and insightful for a comprehensive geometric description. This generalizes the notion of covariance to the unconstrained space of fields, thereby facilitating the description of geometry through globally-defined tensorial quantities.

Furthermore, working directly with the recursion relation does not rely on the existence of an ``all-knowing'' metric. In our approach, amplitudes are not explicitly related to the metric, the corresponding Levi-Civita connection, or the Riemann curvature. Instead, they are calculable from the related n-point correlation functions. Being geometric from our perspective means relating off-shell amplitudes to tensors that transform covariantly under general field redefinitions. In fact, the idea of relegating a metric to a secondary role is not new. For example, in \cite{Cohen:2025prs}, the authors pointed out that it is hard to extract a unique metric for a given Lagrangian, and there are different metric choices which define different Riemannian curvatures even if they ultimately lead to the same EFT. Moreover, it was shown that for a general scalar EFT action, there always exists a metric for which the curvature is identically zero. More discussion will be given on this point later in Section \ref{sec: geometric interpretation}. 

Our approach is based on the following three building blocks: an underlying manifold, a scalar on the manifold to serve as the generating functional, and a connection to impose covariance. We work with a functional manifold, and the fundamental scalar term we choose is the effective action defined in configuration space. In contrast to the approach in \cite{Cohen:2022uuw, Cohen:2023ekv}, where the ``on-shell" connection is defined in terms of 2-point and 3-point functions, we take a step forward by introducing a rank $(0,2)$ tensor\footnote{Under a field redefinition, a $(0,2)$ tensor is an object that transforms like $g_{ab}$ in \eqref{eq: g transform}.}  and its corresponding Christoffel symbols. We define a set of modified $n$-point correlation functions $\mathcal{K}_n$ such that $\mathcal K_{n+1}=\nabla_{n+1}\mathcal K_{n}$, where $\nabla$ is the associated covariant derivative. Then we show that these modified correlation functions reduce to the usual correlation functions when on-shell conditions are imposed, thus producing the same on-shell amplitudes. It follows that the previous equation is naturally recognized as the covariant off-shell recursion relation. Crucially, we must work with  massless theories (scalar theories, here) in order for the earlier reduction to work.

We are aware that working with an infinite-dimensional manifold leads to some unsolved subtleties, such as the missing rigorous definitions of curvature, torsion, etc. That being said, we admit our ignorance and take a modest approach: we follow the conventional definition of geometric quantities in the Riemannian case and replace partial derivatives with functional derivatives. A more rigorous derivation will be given in a follow-up paper. Throughout this paper as a first example, we  limit ourselves to the most general theory with only (massless) scalars at tree-level for simplicity, where the effective action $\Gamma[\phi]$ is just the classical action $S[\phi]$, i.e.  $\Gamma[\phi]=S[\phi]=\int d^4x \mathcal L(\phi)$. We leave cases with fermions and gauge bosons for future studies. Loop-level calculations and direct supersymmetric extensions also seem plausible without conceptual challenges but are beyond the scope of this paper. 

The rest of this paper is organized as follows. In Section \ref{sec:recur} we review the formulation of $n$-pt functions and off-shell recursion relations and their interpretations in a functional geometry, following the convention and derivation in \cite{Cohen:2023ekv}. We proceed in Section \ref{sec: covariant functional manifolds} by showing that the ``on-shell'' covariant recursion relation strongly indicates an ``off-shell'' extension, which is realized by introducing Christoffel symbols to eliminate all terms that vanish on-shell. The above construction works only if we can extract a prior $(0,2)$ tensor from the theory, and we give one of such choices based on geometry-kinematics duality. In Section \ref{sec: good connection} we introduce a ``true'' connection, which is essential in defining the modified $n$-point correlation functions $\mathcal K_n$ that lead to the off-shell recursion relation $\mathcal K_{n+1}=\nabla_{n+1}\mathcal K_{n}$. Specifically, we work through the case where $n=4$ in Section \ref{sec:onshell} by taking the on-shell limit to prove $\mathcal K_n\xrightarrow{on-shell}\mathcal M_n$, and then generalize to the arbitrary case in Section \ref{sec:induction} by induction. The potential caveats and necessary conditions for the covariant framework to work -- why we need massless fields -- are discussed in Section \ref{sec: necessary conditions}. We then discuss the geometric interpretation of this formulation in Section \ref{sec: geometric interpretation} and possible extensions in Section \ref{sec: future}. We review the implementation of on-shell conditions in Appendix \ref{sec: on-shell conditions}. Appendix \ref{sec: appendix} presents an explicit derivation of $\mathcal M, \mathcal N$ and $\mathcal K$ for an toy theory with a $\phi (\partial_\mu \phi)(\partial^\mu \phi)$ interaction. Finally, in Appendix \ref{sec: massive} shows more details of how our approach falls short in theories with massive fields.

\section{Recursion relations as functional geometry}\label{sec:recur}

\begin{figure}
\includegraphics[width=15cm]{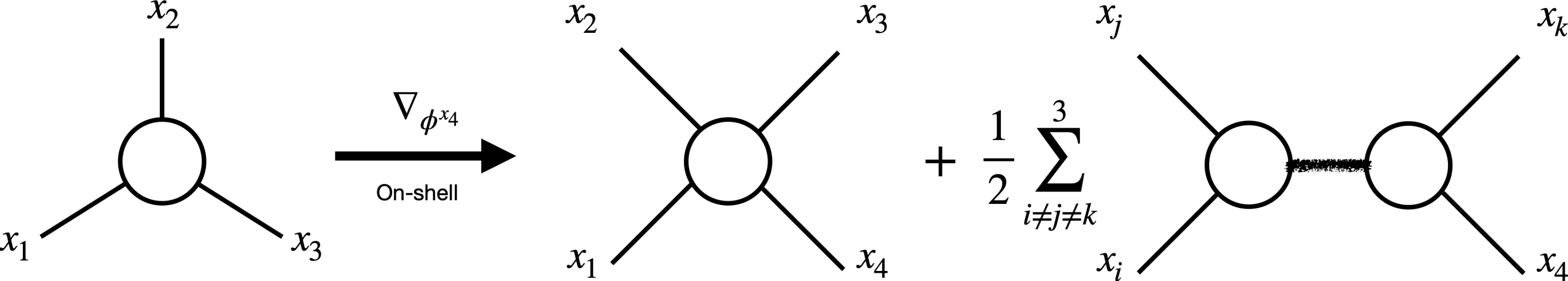}
\caption{Diagrammatic evaluation of Eq.~\eqref{eq:recursion}}
\label{figure1}
\end{figure}

 The starting point of the functional geometry~\cite{Cohen:2022uuw, Cohen:2023ekv} formulation of a QFT is the off-shell recursion relation \cite{Berends:1987me} of source-dependent, amputated correlation functions $\mathcal M_{x_1\cdots x_n}$. These objects are functions of the spacetime locations $x_i$. They can be transformed into amplitudes via the LSZ reduction~\footnote{Specifically, by taking all sources to zero, including any  field strength residue factors, then Fourier transforming $x_i \to p_i$ and evaluating at the $p^2_i = 0$.}. As shown in Ref.~\cite{Cohen:2022uuw, Cohen:2023ekv}, these correlation functions can be massaged into the form
\begin{align}
    \mathcal{M}_{x_1 \cdots x_n x_{n+1}}=\frac{\delta}{\delta \phi^{x_{n+1}}} \mathcal{M}_{x_1 \cdots x_n}-\sum_{i=1}^n G_{x_{n+1} x_i}^y \mathcal{M}_{x_1 \cdots \hat{x}_i y \cdots x_n}.
    \label{eq:recursion}
\end{align}
Here, $\mathcal M_{x_1 \cdots x_{n+1}}$ is the $n+1$ point amputated correlation function and 
\begin{align}
    G_{x_1 x_2}^y \equiv  \mathcal{M}_{x_1 x_2 z} (\mathcal M^{-1})^{z y},
    \label{eq:old G connection}
\end{align} 
where $(\mathcal M^{-1})^{z y}$ is the inverse two-point function (the propagator, up to a factor of $i$) and  $\mathcal{M}_{x_1x_2z}$ is the three-point function\footnote{Since we are defining our n-point functions \emph{without} the factor of $i$ our signs and normalizations differ from other choices in the literature.}. Both of these are determined by functional derivatives of the effective action $\Gamma[\phi]$ with respect to the field $\phi$
\begin{equation}
\begin{split}
\mathcal M_{x_1x_2z}&=-\frac{\delta^3\Gamma}{\delta\phi^{x_1}\delta\phi^{x_2}\delta\phi^{z}},\\
(\mathcal M^{-1})^{zy}&=-\left(\frac{\delta^2\Gamma}{\delta\phi^{z}\delta\phi^{y}}\right)^{-1}.
\end{split}
\end{equation}
Here, and in Eq.~\eqref{eq:recursion}, we use a notation where superscripts and subscripts indicate spacetime arguments, so $\phi^{x_i} = \phi(x_i)$. As a result, $\frac{\delta}{\delta\phi^x}$ denotes the functional derivative with respect to the field $\phi$ evaluated at the space-time point $x$; we will adopt this notation throughout this paper. Finally, for use later on, we distinguish between the indices $x_i$ in Eq.~\eqref{eq:recursion} and the index $y$. The former are `external' in that they are the spacetime arguments of external legs in Fig.~\ref{figure1}, while the latter are `internal'. This will be important as the momenta for external legs are set to specific values when taking the on-shell limit.

What Eq.~\eqref{eq:recursion} says in words (see Fig. \ref{figure1}) is that higher $(n>3)$  point correlation functions are the sum of new contact terms -- coming from additional functional derivatives  -- plus the result from taking correlation functions with one fewer leg and turning one of the legs (eg. $x_i$) into two legs by attaching $x_i$ to a three-point vertex via a propagator. We carry out the stitching on each leg of $\mathcal M_n$, hence the sum over $n$. The hat on $\hat x_i$ in Eq.~\eqref{eq:recursion} means that the index is removed and has been replaced with $y$, e.g. $\mathcal M_{x_1x_2\hat x_3,y x_4} = \mathcal M_{x_1x_2yx_4}$. Said another way, $y$ is the leg that is converted into two external legs via the combination of a propagator and the three-point vertex. 

Our goal is to interpret this recursion relation geometrically, which means that we want to combine the action of $\frac{\delta}{\delta \phi^{x_{n+1}}}$ and $G^y_{x_{n+1} x_i}$ into a covariant derivative $\nabla_{x_{n+1}}$. In this perspective, the objects acted on by this covariant derivative -- the $n$-point correlation functions -- live in a manifold that combines field space with configuration space. The basis vectors of this manifold are functional derivatives, which measure how quantities change as we vary $\phi$ or any of its (partial) derivatives. The connection on this manifold is $G^y_{x_{n+1}x_n}$, the combination of the propagator and the three-point vertex, and the recursion relation above would become
\begin{align}
\label{eq:georecur}
\mathcal M_{x_1 \cdots x_nx_{n+1}} = \nabla_{n+1}\mathcal M_{x_1 \cdots x_n}.
\end{align}
With a fully geometric approach to correlation functions, one would hope that all physical quantities (S-matrix elements) could be expressed as geometric invariants, making it easy to see that they are unchanged under field redefinitions $\phi \to \phi[\phi']\equiv \phi(\phi',\partial\phi',\cdots)$.

As mentioned earlier, the functional interpretation is significantly more ambitious than a `field-space' geometry \cite{Alonso:2015fsp, Alonso:2016oah}; it incorporates derivative redefinitions by working with functional derivatives, extending the field-space approach that merely deals with non-derivative field redefinitions. Furthermore, it is expected that the functional approach reduces to the field-space manifold when we eliminate higher derivatives on fields. More details are provided in Section \ref{sec: geometric interpretation}.

Unfortunately, the functional geometry interpretation of Eq.~\eqref{eq:recursion} does not work. As shown in Ref.~\cite{Cohen:2022uuw, Cohen:2023ekv}, the n-point functions  are not covariant. Under a field redefinition $\phi[\phi']$,
\begin{align}
\label{eq:onshellcov}
    {\mathcal{M}'}_{x_1 \cdots x_n}=\frac{\delta \phi^{y_1}}{\delta {\phi'}^{x_1}} \cdots \frac{\delta \phi^{y_n}}{\delta{\phi'}^{x_n}} \mathcal{M}_{y_1 \cdots y_n}+U_{x_1 \cdots x_n}
\end{align}
The $U_{x_1 \cdots x_n}$ are called ``anholonomic" terms and spoil covariance. While the full, off-shell $\mathcal M_{x_1 \cdots x_n}$ are not covariant, the correlation functions are covariant if we restrict them to be {\it on-shell}, where $U_{x_1\cdots x_n}|_{on-shell}=0$. In other words, if we denote on-shell quantities with a tilde, 
\begin{align}\label{eq:onshellcov2}
    \widetilde{{\mathcal{M}'}}_{x_1 \cdots x_n}=\frac{\delta \phi^{y_1}}{\delta {\phi'}^{x_1}} \cdots \frac{\delta \phi^{y_n}}{\delta {\phi'}^{x_n}} \widetilde{\mathcal{M}}_{y_1 \cdots y_n},
\end{align}
which we will call ``on-shell covariance''. Note that the on-shell external wave function transforms as $\phi'^x(\bar p)=(\frac{\delta\phi'^x}{\delta\phi^y}|_{\phi=\phi_v})\phi^y(\bar p)$, where $\bar p$ is the on-shell momentum. The transformation properties of the on-shell covariant amputated correlation functions \eqref{eq:onshellcov2} exactly cancel with the transformation properties of the external wave functions to yield invariant amplitudes \cite{Cohen:2023ekv}. 

Before continuing, let us spell out exactly what we mean by going `on-shell'. We use the same criteria as Ref.~\cite{Cohen:2023ekv, Cohen:2022uuw}, namely we: i.) evaluate the fields at the physical vacuum ($\phi_v$ in Ref.~\cite{Cohen:2023ekv}), ii.) Fourier transform all correlation functions to momentum space, and iii.) evaluate them at momenta satisfying $p^2_i = 0$ where $p_i$ is the momentum of a massless external particle $i$. 

Let's see how this works for the simplest objects (smallest correlation functions). We define the `one-point' function $\mathcal M_x\equiv \frac{\delta\Gamma[\phi]}{\delta\phi^x}$, where $\Gamma$ is the effective action (See Appendix~\ref{sec: on-shell conditions} for more details). However, the first functional derivative of the effective action is just $J_x$, the source for $\phi(x)$. The physical vacuum of the theory is defined as $\phi = \phi_v$ such that $J_x = 0$ -- the ``zero source condition". Therefore, as $\mathcal M_x = J_x$, $\widetilde{\mathcal M}_x = 0$ regardless of whether or not $\phi^x$ corresponds to an external leg.

The second functional derivative of the effective action, defined as ${\mathcal M}_{x y}\equiv \frac{\delta^2\Gamma[\phi]}{\delta \phi^x\delta\phi^y}$ is related to the inverse of the propagator $D(x,y)$:
\begin{equation}\label{eq:propa}
    D^{-1}(x,y)=-\frac{\delta^2(\Gamma[\phi])}{\delta \phi^x\delta\phi^y}.
\end{equation}
On shell, we evaluate this object at the true vacuum, and then Fourier transform:
\begin{align}
\widetilde{\mathcal M}_{xy} \equiv \int d^4x e^{i\,p\,x}\frac{\delta^2(\Gamma[\phi])}{\delta \phi^x\delta\phi^y}\Big|_{\phi = \phi_v}.
\end{align}
If the momentum $p$ satisfies $p^2 = m^2$, this vanishes. Thus, on shell we have 
\begin{equation}
\label{eq:onshell12}
    \widetilde{\mathcal M}_x=0,\quad \widetilde{\mathcal M}_{x y}=0.
\end{equation}
Importantly, the zero source condition applies regardless of the argument of $\phi$, as it defines the ground state of the physical configuration, while the two-point function vanishes only when contracted with the external wavefunction ($e^{ipx}$ in the above equation). This translates to taking the on-shell condition for {\it at least one} of the indices of $\widetilde{\mathcal M}_{xy}$. Performing the same steps, we can take the on-shell limit of higher point $\mathcal M$. For most purposes this boils down to just imposing Eq.~\eqref{eq:onshell12}, however there is an important caveat: in higher point $\mathcal M$, we must be on the lookout for terms that have poles in the on-shell limit, as these will complicate or nullify the limits in Eq.~\eqref{eq:onshell12}. We will say more about this caveat in the next section, and explore what restrictions this places on scalar theory parameters in more detail in Section \ref{sec: necessary conditions}.

The goal of this paper is to adjust the functional geometry formalism such that the recursion relation also holds off-shell. We want to continue working on a field space plus configuration space manifold, so we will continue to use functional derivatives as our basis vectors. Therefore, all adjustments to the above formalism must involve adjustments to $\mathcal M$. As $G$ is just a combination of two- and three-point $\mathcal M$, adjustments in $\mathcal M$ will imply adjustments to $G$, and as we want to maintain on-shell covariance and reduce to the on-shell version of Eq.\eqref{eq:recursion}, all adjustments that we make to $\mathcal M$ must vanish on-shell.

 \section{A covariant functional manifold}\label{sec: covariant functional manifolds}

The non-covariant properties of $\mathcal M$ are apparent even in the two-point function, 
\begin{align}
\mathcal{M}_{x_1x_2}\rightarrow{{\mathcal{M}'}_{x_1x_2}}=\frac{\delta\phi^{y_1}}{\delta{\phi'}^{x_1}}\frac{\delta\phi^{y_2}}{\delta{\phi'}^{x_2}}\mathcal{M}_{y_1y_2}+\frac{\delta^2\phi^y}{\delta{\phi'}^{x_1}\delta{\phi'}^{x_2}}\mathcal{M}_y.
\end{align}
We can fix this issue if our setup contains a genuine rank $(0,2)$ tensor. Lets suppose this $(0,2)$ tensor exists and it is called  $T(\phi)_{x_1x_2}$ where, under $\phi \to \phi'[\phi]$ 
\begin{align}
    T_{x_1 x_2}(\phi) \rightarrow {T'}_{x_1 x_2}({\phi}')=\frac{\delta{\phi}^{y_1}}{\delta \phi'^{x_1}} \frac{\delta{\phi}^{y_2}}{\delta \phi'^{x_2}} T_{y_1 y_2}(\phi).
\end{align}
Using $T_{x_1x_2}$, we can build a Christoffel symbol
\begin{align}\label{eq: T connection}
\mathbf{\Gamma^y_{x_1x_2}} = \frac{1}{2} T^{y z}\left(T_{x_1 z,x_2}+ T_{x_2 z,x_1}- T_{x_1 x_2,z}\right),
\end{align}
where the ``$,$" indicates a functional derivative, e.g. $T_{x_1x_2,z} = \frac{\delta}{\delta \phi^z}T_{x_1x_2}$. With this, the combination
\begin{align}
\label{eq:Nabdefn}
    \mathcal N_{x_1x_2} \equiv \mathcal M_{x_1x_2} - \mathbf{\Gamma^{y}_{x_1x_2}}\mathcal M_y
\end{align}
is now covariant, $\mathcal N'_{x_1x_2} = \frac{\delta \phi^{y_1}}{\delta \phi'^{x_1}}\frac{\delta \phi^{y_2}}{\delta \phi'^{x_2}} \mathcal N_{y_1y_2}$, as is the three-point function 
\begin{align}\label{eq:Nabc}
    \mathcal N_{x_1x_2x_3} = \mathcal M_{x_1x_2x_3} - \mathbf{\Gamma^y_{x_2\,x_3}}\mathcal M_{y\,x_1} - \mathbf{\Gamma^y_{x_1\, x_3}} \mathcal M_{y\,x_2} - \mathbf{\Gamma^y_{x_1x_2}}\mathcal M_{y\,x_3} - \mathbf{\Gamma^{y}_{x_1\,x_2,x_3}}\mathcal M_y.
\end{align}

At this point, we emphasize two things. First, $\mathcal N$ and $\mathcal M$ are not equivalent. So, for this line of reasoning to be at all useful, we need $\mathcal N \to \mathcal M$ on-shell. Second, what is this rank $(0,2)$ tensor, and how can we be sure that such an object is at our disposal in any given (scalar) theory?  

To answer the second question, we need to look no further than $g_{ij}$, the prefactor of the two-derivative Lagrangian term $\partial_\mu \phi_i \partial^\mu \phi_j$. As established in a setup called geometric-kinematic duality~\cite{Cheung:2022vnd}, $g_{ij}$ transforms as a $(0,2)$ tensor under generic field redefinitions in momentum space by mapping the kinematic information onto the NLSM and making use of the results from field-space geometry. In addition, higher derivative terms, potential terms, etc. can be squeezed into $g_{ij}$, all while maintaining their covariance even under derivative field redefinitions. As a consequence, the kinetic metric $g_{ij}$ is a good candidate serving as the $(0,2)$ tensor. However, we should emphasize that this choice is not unique -- an aspect that prevents us from treating $g_{ij}$ as the metric in our construction.\footnote{$g_{ij}$ does transform as a metric, but it is not ``the metric'' in our formulation, as our result is independent of the induced curvature, etc., due to the non-uniqueness.} To see the non-uniqueness, let us shift $g_{ij} \to g_{ij} + h_{ij}$. This will change the tensor itself, as it shifts the coefficient of the two-derivative term. However, this shift will not affect any physics if $h_{ij}$ satisfies (going to momentum space for convenience)
\begin{align}
    \int \frac{d^4p}{(2\pi)^4}\frac{d^4q}{(2\pi)^4}h_{ij}(p,q)\, p \cdot q\, \phi^i(p)\phi^j(q) = 0.
\end{align}

From our perspective, $g_{ij}$ is merely a $(0,2)$ tensor, so the non-uniqueness is not a roadblock. We will later determine what the metric and connection are on our manifold.

This attitude of the role of the metric, is similar to the perspective in Ref.~\cite{Cohen:2025prs}. There, it was argued that the connection is the key object needed to make amplitudes covariant; the metric comes second, and is most useful when one wants to relate functional geometry results to field space geometry. In particular, the ambiguity in the metric of functional geometry translates to an ambiguity in the basis to write operators in the Lagrangian. To most cleanly connect with results from field geometry, Ref.~\cite{Cohen:2025prs} proposes a ``Warsaw'' metric that corresponds exactly to the coefficient of the two-derivative Lagrangian term.

Before moving on, now that we have an example of what can serve as a $(0,2)$ tensor $T_{xy}\equiv g_{xy}$, let us verify that the two and three point correlation functions $\mathcal N$ and $\mathcal M$ are equivalent on-shell. For the two point function, $\widetilde{\mathcal N}_{x_1x_2}$ and $\widetilde{\mathcal M}_{x_1x_2}$ differ by a term $\propto \widetilde{\mathcal M}_y$, which vanishes under our conditions Eq.~\eqref{eq:onshell12} regardless of whether or not $y$ is an external index. The three point functions differ by more terms, but from Eq.~\eqref{eq:Nabc} we see all extra terms involve either $\widetilde{\mathcal M}_{yx_i}$, where $x_i$ is an external index, or $\widetilde{\mathcal M}_y$. Both of these objects vanish in the on-shell limit following ~\eqref{eq:onshell12} , so:
\begin{align}
\label{eq:basiconshell}
\widetilde{\mathcal N}_{x_1x_2} &= \widetilde{\mathcal M}_{x_1x_2} \nonumber \\
\widetilde{\mathcal N}_{x_1x_2x_3} &= \widetilde{\mathcal M}_{x_1x_2x_3}
\end{align}
Higher-point functions $\mathcal{N}_{x_1x_2\cdots x_n}$ can be constructed recursively using $\mathbf{\Gamma^c_{a b}}$, and they are essential for building on-shell recursion relations. We will briefly discuss this point in the next section, and more details can be found in \cite{Cohen:2024bml, Cohen:2025prs}. However, these $\mathcal N$ functions are insufficient for the purposes of this paper, which primarily focuses on the off-shell construction, and we need additional structures to be introduced shortly.

Before moving on, we must raise an important, but subtle caveat. In deriving Eq.~\eqref{eq:basiconshell} we have assumed that $\mathbf{\Gamma}$ is well behaved, meaning it does not cancel $\widetilde{\mathcal M}_{xy} \to 0$. This could happen if $\widetilde{\mathbf{\Gamma}}$ has a pole for the same momentum configuration where $\widetilde{\mathcal M}_{xy}$ is zero, e.g.
\begin{align}
\mathcal M_{xy} \sim p^2_x,\quad \mathbf{\Gamma^x_{x_1x_2}} \sim \frac{1}{p^2_x},
\end{align}
where we have worked in momentum space.
The result of this cancellation, $\widetilde{\mathbf{\Gamma}}^y_{x_1x_2}\widetilde{\mathcal M}_{yx_3} \ne 0$ ($x_3$ an external leg), disrupts the equality of $\mathcal N$ and $\mathcal M$ on-shell and spoils our approach. For the next few sections, we will assume that all combinations $\widetilde{\mathbf{\Gamma}}^y_{x_1x_2}\widetilde{\mathcal M}_{yx_3} = 0$, returning to a detailed study of the restrictions that places on our theory in Sec.~\ref{sec: necessary conditions}.

\section{A ``good'' connection}\label{sec: good connection}

With the introduction of a $(0,2)$ tensor $T$ and the Christoffel symbols $\mathbf{\Gamma}$ (and caveat), we've found a quantity, $\mathcal N$, that is covariant, even off-shell.\footnote{In the previous section, we identified that $g_{ij}$ can serve as this type of object, but to emphasize that it is not the only possibility we will use $T$.} However, these $\mathcal N$ do not satisfy the recursion relation Eq.~\eqref{eq:georecur}, in the sense that they do not contain the same information as an actual $n$-point function. This makes some intuitive sense, as the $\mathcal M$ are derived from the whole effective action, while the related Christoffel symbols $\mathbf{\Gamma}$, i.e. 3-point vertices, are read off of the two-derivative term in the (tree-level) Lagrangian. 

To be more explicit, it is straightforward to see that $\mathbf{\widetilde \Gamma^y_{x_1x_2}}\neq G^y_{x_1x_2}$ (cf. Eq.~\ref{eq:recursion}) when evaluated on-shell, and therefore the equation 
\begin{align}
\mathcal N_{x_1x_2\cdots x_n} = (\mathcal{M}_{x_1x_2\cdots x_{n-1},x_n}-\sum_{i=1}^{n-1} \mathbf{\Gamma^y_{x_nx_i}}\mathcal{M}_{x_1\cdots \hat{x_i}y\cdots x_{n-1}})
\end{align}
does not reproduce the correct $n$-point amplitudes.

 To relate the Christoffel symbols to quantities defined by the effective action, and thereby determine the true connection on our manifold, we write
\begin{align}
\label{eq:Gammadef}
    \Gamma_{x_1 x_2}^y \equiv \frac{1}{2}\left(\mathcal{N}^{-1}\right)^{y z}\left(\mathcal{N}_{x_1 z , x_2}+\mathcal{N}_{x_2 z , x_1}-\mathcal{N}_{x_1 x_2 , z}\right),
\end{align}
where, as before, the ``," indicates functional differentiation. Combing this connection with the functional derivative, we form the true covariant derivative $\nabla_{x_n}$. As we will verify shortly, this covariant derivative acts on quantities $\mathcal K_{x_1 \cdots x_n}$ which are related to the $n$-point correlation functions $\mathcal M$ (and $\mathcal N$),
\begin{align}
\label{eq:offshellrecur}
 \mathcal K_{x_1 \cdots x_n x_{n+1}} &= \nabla_{x_{n+1}}\mathcal K_{x_1 \cdots x_n} = \mathcal K_{x_1 \cdots x_n, x_{n+1}} - \sum\limits_{i=1}^{n}\Gamma^y_{x_{n+1} x_i} \mathcal K_{x_1 \cdots \hat x_i\,y \cdots x_n}
\end{align}
Defining $\mathcal K_{x_1x_2} = \nabla_{x_2}\mathcal M_{x_1} = \mathcal M_{x_1,x_2} - \Gamma^y_{x_2 x_1}\mathcal M_y$, we find the recursion relation:
\begin{align}\label{eq:recursion k}
\mathcal K_{x_1 \cdots x_n x_{n+1}} = \nabla_{x_{n+1}}\nabla_{x_n}\cdots \nabla_{x_2}\mathcal M_{x_1}.
\end{align}
The $\mathcal K$ are covariant by construction, even off-shell and  when considering derivative field redefinitions, and reduce to $\mathcal M$'s when on-shell. 

Before proving Eq.~\eqref{eq:offshellrecur} and establishing that it has the correct on-shell limit, it is worth comparing this approach with the approach in Ref.~\cite{Cohen:2024bml, Cohen:2025prs}. In~\cite{Cohen:2024bml}, the authors define $\mathcal M_{x_1\cdots x_n}$ ($S_{,x_1\cdots x_n}$ in their convention, with the "$,$" indicating functional derivative) and improve it to an off-shell covariant quantity by utilizing Christoffel symbols derived from $g_{ij}$. This gives the quantity $\mathcal N_{x_1 \cdots x_n}$ in our notation, or $S_{;x_1\cdots x_n}$ in their notation (the ";" now indicating the derivatives have been improved by combining the functional derivative with the Christoffel symbols from $T_{ij}\equiv g_{ij}$). 

The next step is where the two approaches differ. Reference~\cite{Cohen:2024bml} uses the $\mathcal N_{x_1\cdots x_n}$ to write correlation functions as a sum of {\it individually on-shell} covariant pieces. The propagator -- which enters Eq.~\eqref{eq:recursion} via $G$ -- is on-shell covariant, so the changes taking $\mathcal M \to N$ ($S_{,} \to S_{;})$ can be phrased as new vertex Feynman rules. The $k$-particle vertex $\mathcal V_{1\cdots k}$ is defined as \begin{align}\label{eq: on-shell covariant v}
    \mathcal{V}_{1 \cdots k} \equiv \mathcal N_{1 \cdots k}+\mathbf{\Gamma_{1 \cdots k}}^a \mathcal N_{, a}+\sum_{\hat b \in \text { external }} \mathbf{\Gamma_{1 \cdots \hat b \cdots k}}^a \mathcal N_{, a b}
\end{align}
where $1,2,\cdots,k$ are external indices and $a$ is the internal index; the $\mathbf{\Gamma}$ in the last term on the right-hand side are generalized Christoffel symbols and $\hat b$ means that the index $b$ is absent in $\mathbf{\Gamma}$, see~\cite{Cohen:2024bml}. The $\mathcal N_{1 \cdots k}$ are off-shell covariant by definition, while $\widetilde{\mathcal N}_a = \widetilde{\mathcal M}_a$ and $\widetilde{\mathcal N}_{ab} = \widetilde{\mathcal M}_{ab}$ ($b \in $ external legs) both vanish on-shell. Thus, the $\mathcal V_{1\cdots k}$ are on-shell covariant. 

\begin{figure}[t]
\includegraphics[width=15cm]{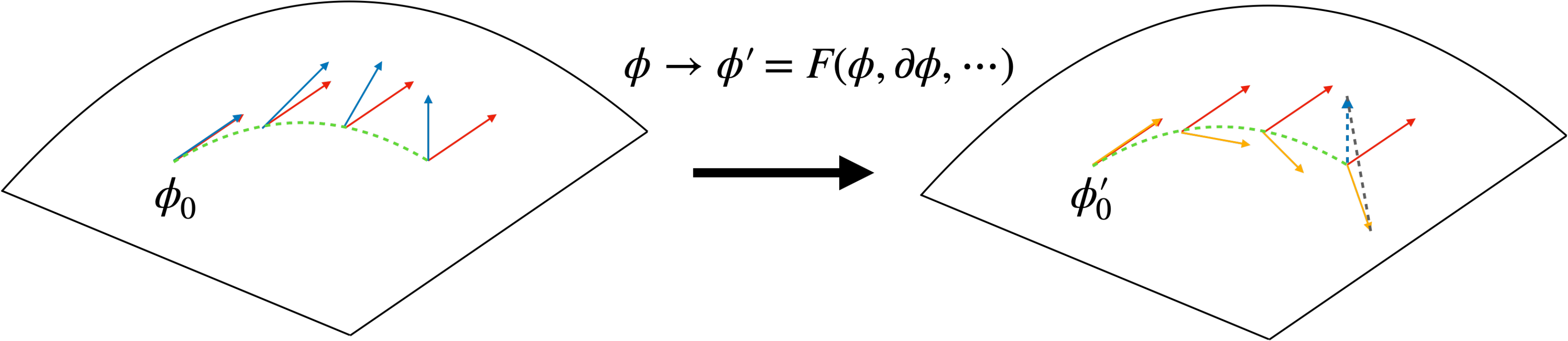}
\caption{$\mathcal{K}$ (defined in Eq. \eqref{eq:offshellrecur}) is represented by the red vectors which remain covariant after the field redefinitions since they are ``real'' tensors defined on the functional manifold. In contrast, $\mathcal{M}$ is represented by the blue and orange vectors, which remain covariant only if evaluated at the on-shell points, denoted $\phi_0$ and $\phi'_0$ in the figure. The difference (dashed line) is exactly the contribution from the ``anholonomic'' terms.}
\label{figure2}
\end{figure}

In our approach, we take the $\mathcal N$ then define a new quantity $\mathcal K$. The $\mathcal K$ are off-shell covariant and obey a recursion relationship. They are different from $\mathcal M$, but by amounts which vanish on-shell. An illustration of  how $\mathcal M$ and $\mathcal K$ transform under a general field redefinition is shown in Fig. \ref{figure2}. The main difference is that $\mathcal M$ is covariant only at the vacuum point $\phi_0$, i.e. satisfying Eq. \eqref{eq:onshellcov2}, where the `anholonomic' terms vanish; while $\mathcal K$ is covariant even when evaluated off-shell (without the tilde):
\begin{align}\label{eq:onshellcovk}
    {{\mathcal{K}'}}_{x_1 \cdots x_n}=\frac{\delta \phi^{y_1}}{\delta {\phi'}^{x_1}} \cdots \frac{\delta \phi^{y_n}}{\delta {\phi'}^{x_n}} {\mathcal{K}}_{y_1 \cdots y_n},
\end{align}
as a direct consequence of Eq. \eqref{eq:recursion k}.

Note that, in the approach of Ref.~\cite{Cohen:2024bml} there is no pole subtlety in the on-shell covariant framework regardless of whether the theory being massive or massless. What is important is that Eq.\eqref{eq: on-shell covariant v} transforms covariantly\footnote{See App. A of \cite{Cohen:2025prs} for more details.}, not whether or not $\widetilde{\mathbf{\Gamma}}^y_{x_1x_2}\widetilde{\mathcal M}_{yx_3} = 0$. In contrast, in our off-shell approach $\widetilde{\mathbf{\Gamma}}^y_{x_1x_2}\widetilde{\mathcal M}_{yx_3} = 0$ is a necessary and sufficient condition, the implications of which will be further explored in Sec.~\ref{sec: necessary conditions}.

\subsection{On-shell limit}\label{sec:onshell}

The next step is to show that this function reduces to $\mathcal{M}_{x_1x_2\cdots x_n}$ when evaluated at the on-shell point, i.e. 
\begin{equation}\label{eq:onshell equiv}
    \widetilde{\mathcal{K}}_{x_1x_2\cdots x_n}=\widetilde{\mathcal{M}}_{x_1x_2\cdots x_n}\equiv(\mathcal{M}_{x_1x_2\cdots x_{n-1},x_n}-\sum_{i=1}^{n-1} G^y_{x_nx_i}\mathcal{M}_{x_1\cdots \hat{x_i}y\cdots x_{n-1}})|_{on-shell}
\end{equation}
where the ``," indicates functional differentiation and $\Gamma$ is the connection formed from $\mathcal N$ defined in Eq.~\eqref{eq:Gammadef}. 

The two and three point cases are trivial, meaning they don't even require evaluating $\Gamma$ on-shell, $\widetilde{\mathcal K}_{x_1x_2} = \widetilde{\mathcal M}_{x_1x_2} $, $\widetilde{\mathcal K}_{x_1x_2x_3} = \widetilde{\mathcal M}_{x_1x_2x_3}$ by the same logic used in Eq.~\eqref{eq:basiconshell} -- and with all indices corresponding to external legs, just with the `genuine' connection $\Gamma$ replacing $\mathbf{\Gamma}$. For readers seeking a more concrete example, Appendix \ref{sec: appendix} provides an illustration of this statement, where we work with an explicit derivative field redefinition and calculate the corresponding 2-point $\mathcal M_{xy},\mathcal N_{xy}$ and $\mathcal K_{xy}$ and their on-shell limits.

The first non-trivial check comes when looking at the on-shell limit of $\Gamma$ itself. From Eq.~\eqref{eq:Gammadef}, we see that $\Gamma^y_{x_1x_2}$ involves $\mathcal N_{xy,z}$, the functional derivative. Applying functional derivative to the pieces in Eq.~\eqref{eq:Nabdefn}, we get
\begin{align}
\mathcal N_{xy,z} = \mathcal M_{xyz} - \mathbf{\Gamma^w_{xy}}\mathcal M_{wz} - \mathbf{\Gamma^{w}_{xy,z}} \mathcal M_w.
\end{align}
Taking the on-shell limit, the last term vanishes by the zero source condition (Eq. \ref{eq:onshell12}), but the fate of the middle term depends on whether or not the index $z$ corresponds to an external leg. If $z$ is an external leg index, then $\widetilde{\mathcal M}_{wz}$ vanishes on-shell regardless of whether $w$ is an external leg and $\widetilde{\mathcal N}_{xy,z} = \widetilde{\mathcal M}_{xyz}$. However, if $z$ is not an external leg index (so that neither index of $\mathcal M_{wz}$ corresponds to an external leg), the on-shell condition does not apply and $\mathcal M_{wz}$ is not guaranteed to vanish -- even with our caveat that $\mathbf{\widetilde{\Gamma}}$ is well-behaved. 

Applied to $\Gamma^y_{x_1x_2}$, where $\{x_1, x_2\}$ are external legs,
\begin{align}
\label{eq:onshellconnect}
\widetilde{\Gamma}^y_{x_1x_2} \equiv \frac 1 2 (\widetilde{\mathcal M}^{-1})^{yz}(\widetilde{\mathcal M}_{x_1 z x_2} + \widetilde{\mathcal M}_{x_2 z x_1} - \widetilde{\mathcal N}_{x_1x_2,z}),
\end{align}
where we use the Woodbury identity $(A+B)^{-1}=A^{-1}-A^{-1}B(A+B)^{-1}$ to handle $(\mathcal N^{-1})$~\footnote{Specifically, $\mathcal N_{xz} = \mathcal M_{xz} - \Gamma^y_{xz}\mathcal{M}_y$, so $(\mathcal N^{-1})_{xz} = (\mathcal M^{-1})_{xz} - (\mathcal M^{-1})_{xw} \Gamma^q_{wr}\mathcal M_{q}(\mathcal M_{rz} - \Gamma^q_{rz}\mathcal M_q)^{-1}$, where the second piece vanishes on-shell by the zero source condition, $\widetilde{\mathcal M}_q = 0$.}. In the first two terms in Eq.~\eqref{eq:onshellconnect}, $\widetilde{\mathcal N} \to \widetilde{\mathcal M}$ on-shell, but last term, $\widetilde{\mathcal N}_{x_1x_2,z}$, does not automatically reduce to $\widetilde{\mathcal M}$ because it contains a factor of $\mathcal M_{wz}$ with neither $w$ nor $z$ an external leg. It seems that we are now facing a problem. However, as $\Gamma$ is defined in terms of $\mathcal N$ -- a symmetric, invertible $(0,2)$ tensor -- we know that the latter is covariantly conserved, $\nabla_z \mathcal N_{x_1x_2} = 0$, where $\nabla$ here is the covariant derivative associated with $\Gamma^y_{x_1x_2}$.\footnote{This condition is called `metric compatibility' in the literature, as the symmetric, invertible $(0,2)$ tensor in General Relativity is the metric $g$. Indeed, we can identify $\mathcal N_{ab}$ as a metric on our manifold (up to subtleties of infinite dimensionality mentioned in the introduction), however we refrain from the terminology to emphasize a broader view of geometry that prioritizes covariant quantities over metric-derived quantities like the curvature.}

 Expanding this out,
\begin{align}\label{eq: compat}
\mathcal N_{x_1x_2,z} &- \Gamma^y_{x_1z}\mathcal N_{yx_2} - \Gamma^y_{x_2z}\mathcal N_{yx_1} = 0 \nonumber \\
\rightarrow\,\, & \widetilde{\mathcal N}_{x_1x_2,z} = \Gamma^y_{x_1z}\widetilde{\mathcal N}_{yx_2} + \Gamma^y_{x_2z}\widetilde{\mathcal N}_{yx_1} = 0
\end{align}
because both $\mathcal N$ factors on the second line contain at least one external index. With $\widetilde{\mathcal N}_{x_1x_2,z}$ dropped, and using the symmetry properties of $\mathcal M_{x_1x_2\cdots}$,
\begin{align}
\label{eq:onshellconnect2}
\widetilde{\Gamma}^y_{x_1x_2} \equiv \frac 1 2 (\widetilde{\mathcal M}^{-1})^{yz}(\widetilde{\mathcal M}_{x_1z, x_2} + \widetilde{\mathcal M}_{x_2z,x_1} ) =  (\widetilde{\mathcal M}^{-1})^{yz}(\widetilde{\mathcal M}_{z x_1 x_2}) = G^y_{x_1x_2},
\end{align}
we see that we recover the desired on-shell connection. 

Now we conclude that the two connections are equivalent when evaluated on-shell and let us calculate the four-point function $\mathcal{K}_{x_1x_2x_3x_4}$,
\begin{equation}\label{eq:4ptK}
\begin{split}
\mathcal{K}_{x_1x_2x_3x_4}&\equiv\nabla_{x_4}\mathcal{K}_{x_1x_2x_3}\\
    &=\mathcal{K}_{x_1x_2x_3,x_4}-\Gamma^y_{x_4x_1}\mathcal{K}_{yx_2x_3}-\Gamma^y_{x_4x_2}\mathcal{K}_{yx_3x_1}-\Gamma^y_{x_4x_3}\mathcal{K}_{yx_1x_2}
\end{split} 
\end{equation}
which we would like to show reduces to 
\begin{align}
\label{eq:M4}
    \widetilde{\mathcal M}_{x_1x_2x_3x_4} \equiv \widetilde{\mathcal M}_{x_1x_2x_3,x_4} - G^y_{x_1 x_4}\widetilde{\mathcal M}_{y x_2 x_3} - G^y_{x_2 x_4}\widetilde{\mathcal M}_{x_1 y x_3} - G^y_{x_3x_4}\widetilde{\mathcal M}_{x_1x_2y}  
\end{align}
 Our strategy will be to isolate the pieces that are proportional to $\mathcal M_y$, $\mathcal M_{x y}$ as these vanish -- up to the subtlety mentioned above if neither $\{x,y\}$ are external indices -- in the on-shell limit. Let's expand the first term in Eq.~\eqref{eq:4ptK}:
\begin{align}
\label{eq:expandK}
\mathcal{K}_{x_1x_2x_3,x_4} &=\frac{\delta}{\delta \phi^{x_4}}\left(\frac{\delta}{\delta \phi^{x_3}}\mathcal{K}_{x_1x_2}-\Gamma^y_{x_1x_3}\mathcal{K}_{yx_2}-\Gamma^y_{x_xx_3}\mathcal{K}_{yx_1}\right) \\
    &=\frac{\delta}{\delta \phi^{x_4}}\left[\frac{\delta}{\delta \phi^{x_3}}(\mathcal{M}_{x_1x_2}-\Gamma^y_{x_1x_2}\mathcal{M}_y)-\Gamma^y_{x_1x_3}(\mathcal{M}_{yx_2}-\Gamma^z_{yx_2}\mathcal{M}_z)-\Gamma^y_{x_2x_3}(\mathcal{M}_{yx_1}-\Gamma^z_{yx_1}\mathcal{M}_z)\right] \nonumber \\
    &=\mathcal{M}_{x_1x_2,x_3x_4}-\Gamma^y_{x_1x_2}\mathcal{M}_{y,x_3x_4}-\Gamma^y_{x_1x_3}\mathcal{M}_{yx_2,x_4}-\Gamma^y_{x_2x_3}\mathcal{M}_{yx_1,x_4}+U_4({\mathcal{M}}_1)+U_4'({\mathcal{M}_2}), \nonumber 
\end{align}
where $U_4({\mathcal{M}}_1)$ and $U_4'({\mathcal{M}_2})$ contain the terms which are proportional to $M_x$ and $M_{xy}$, respectively:
\begin{align}
\label{eq:Uin4pt}
U_4({\mathcal{M}}_1) 
    &\equiv(-\Gamma^z_{x_1x_2,x_3x_4}+\Gamma^y_{x_1x_3,x_4}\Gamma^z_{yx_2}+\Gamma^y_{x_1x_3}\Gamma^z_{yx_2,x_4}+\Gamma^y_{x_2x_3,x_4}\Gamma^z_{yx_1}+\Gamma^y_{x_2x_3}\Gamma^z_{yx_1,x_4})\,\mathcal{M}_z, \nonumber \\
U_4'({\mathcal{M}}_2)
    &\equiv-\Gamma^y_{x_1x_2,x_3}\mathcal{M}_{y,x_4}-\Gamma^y_{x_1x_2,x_4}\mathcal{M}_{y,x_3}-\Gamma^y_{x_1x_3,x_4}\mathcal{M}_{yx_2}-\Gamma^y_{x_2x_3,x_4}\mathcal{M}_{yx_1} \nonumber \\
&\quad\quad\quad\quad+\Gamma^y_{x_1x_3}\Gamma^z_{yx_2}\mathcal{M}_{z,x_4}+\Gamma^y_{x_2x_3}\Gamma^z_{yx_1}\mathcal{M}_{z,x_4}.
\end{align}

The three other terms in \eqref{eq:4ptK} can be also be expanded in terms of $\mathcal{M}$. For example,
\begin{align}
    \Gamma^y_{x_4x_1}\mathcal{K}_{yx_2x_3}= &\, \Gamma^y_{x_4x_1}(\mathcal{K}_{yx_2,x_3}-\Gamma^z_{yx_3}\mathcal{K}_{zx_2}-\Gamma^z_{x_2x_3}\mathcal{K}_{zy})\nonumber \\
    =\, &\Gamma^y_{x_4x_1}\left[\frac{\delta}{\delta\phi^{x_3}}(\mathcal{M}_{yx_2}-\Gamma^z_{yx_2}\mathcal{M}_z)-\Gamma^z_{yx_3}(\mathcal{M}_{zx_2}-\Gamma^w_{zx_2}\mathcal{M}_w)-\Gamma^z_{x_2x_3}(\mathcal{M}_{zy}-\Gamma^w_{zy}\mathcal{M}_w)\right]\nonumber \\
    =\, &\Gamma^y_{x_4x_1}\mathcal{M}_{yx_2,x_3}+V_{41}(\mathcal{M}_1)+V_{41}'(\mathcal{M}_2), 
\end{align}
where $V_{41}({\mathcal{M}}_1)$ and $V_{41}'({\mathcal{M}_2})$ contain the terms which are proportional to $\mathcal{M}_x$ and $\mathcal{M}_{xy}$:
\begin{align}
\label{eq:vdefn}
    V_{41}({\mathcal{M}}_1)\equiv&\,\Gamma^y_{x_4x_1}(-\Gamma^z_{yx_2,x_3}+\Gamma^w_{yx_3}\Gamma^z_{wx_2}+\Gamma^w_{x_2x_3}\Gamma^z_{wy})\mathcal{M}_z, \nonumber \\
    V_{41}'({\mathcal{M}}_2)\equiv&\,-\Gamma^y_{x_4x_1}(\Gamma^z_{yx_2}\mathcal{M}_{z,x_3}+\Gamma^z_{yx_3}\mathcal{M}_{zx_2}+\Gamma^z_{x_2x_3}\mathcal{M}_{zy}).
\end{align}
Notice that $V'$ contains pieces where neither of the indices on $\mathcal M_{yz}$ corresponds to an external leg ($\in \{x_1,x_2,x_3,x_4\}$).

Putting everything together, we can lump together the $U$ and $V$ terms into a single piece $\mathcal E(\mathcal M)$:
\begin{equation}\label{eq:onshell expansion}
    \begin{split}
        \mathcal{K}_{x_1x_2x_3x_4}&=\mathcal{K}_{x_1x_2x_3,x_4}-\Gamma^y_{x_4x_1}\mathcal{K}_{yx_2x_3}-\Gamma^y_{x_4x_2}\mathcal{K}_{yx_3x_1}-\Gamma^y_{x_4x_3}\mathcal{K}_{yx_1x_2}\\
        &=\mathcal{M}_{x_1x_2x_3,x_4}-\Gamma^y_{x_1x_2}\mathcal{M}_{yx_3x_4}-\Gamma^y_{x_1x_3}\mathcal{M}_{yx_2x_4}-\Gamma^y_{x_2x_3}\mathcal{M}_{yx_1x_4}\\
        &-\Gamma^y_{x_4x_1}\mathcal{M}_{yx_2x_3}-\Gamma^y_{x_4x_2}\mathcal{M}_{yx_3x_1}-\Gamma^y_{x_4x_3}\mathcal{M}_{yx_1x_2}-\mathcal{E}_4(\mathcal{M}).
    \end{split}
\end{equation}
Here,
\begin{equation}
    \mathcal{E}_4(\mathcal{M})\equiv U_4(\mathcal{M}_1)+U_4'(\mathcal{M}_2)+\sum_{i=1}^3\left[V_{4i}(\mathcal{M}_1)+V_{4i}'(\mathcal{M}_2)\right],
\end{equation}
where $V_{42}, V_{42}',V_{43}$ and $V_{43}'$ are defined as in Eq.~\eqref{eq:vdefn} but with $x_1 \leftrightarrow x_2, x_1 \leftrightarrow x_3$.

Taking the on-shell limit, all the terms in $\mathcal E(\mathcal M)$ vanish except for the $V'$. Focusing on these terms,
\begin{equation}
\widetilde{\mathcal{E}}_4=\widetilde{V}_{41}'+\widetilde{V}_{42}'+\widetilde{V}_{43}'=-(\widetilde\Gamma^y_{x_4x_1}\widetilde\Gamma^z_{x_2x_3}+\widetilde\Gamma^y_{x_4x_2}\widetilde\Gamma^z_{x_1x_3}+\widetilde\Gamma^y_{x_4x_3}\widetilde\Gamma^z_{x_2x_1})\widetilde{\mathcal M}_{zy}.
\end{equation}
We can now fully take the on-shell limit of \eqref{eq:onshell expansion}:
\begin{equation}
     \begin{split}
        \widetilde{\mathcal{K}}_{x_1x_2x_3x_4}&=( \widetilde{\mathcal{M}}_{x_1x_2x_3;x_4}-\widetilde\Gamma^y_{x_1x_2}\widetilde{\mathcal{M}}_{yx_3x_4}-\widetilde\Gamma^y_{x_1x_3}\widetilde{\mathcal{M}}_{yx_2x_4}-\widetilde\Gamma^y_{x_2x_3}\widetilde{\mathcal{M}}_{yx_1x_4}\\
        &-\widetilde\Gamma^y_{x_4x_1}\widetilde{\mathcal{M}}_{yx_2x_3}-\widetilde\Gamma^y_{x_4x_2}\widetilde{\mathcal{M}}_{yx_3x_1}-\widetilde\Gamma^y_{x_4x_3}\widetilde{\mathcal{M}}_{yx_1x_2})\\
        &+(\widetilde\Gamma^y_{x_4x_1}\widetilde\Gamma^z_{x_2x_3}+\widetilde\Gamma^y_{x_4x_2}\widetilde\Gamma^z_{x_1x_3}+\widetilde\Gamma^y_{x_4x_3}\widetilde\Gamma^z_{x_2x_1})\widetilde{M}_{zy},
    \end{split}
\end{equation}

Comparing this with Eq.~\eqref{eq:M4} and recalling that $\widetilde\Gamma^y_{x_1x_2} = G^y_{x_1x_2}$, we see that in order for Eq.~\eqref{eq:onshell equiv} to hold, we need 
\begin{equation}\label{eq:GammaM equivalence}
(\widetilde\Gamma^y_{x_1x_2}\widetilde{\mathcal{M}}_{yx_3x_4}+\widetilde\Gamma^y_{x_1x_3}\widetilde{\mathcal{M}}_{yx_2x_4}+\widetilde\Gamma^y_{x_2x_3}\widetilde{\mathcal{M}}_{yx_1x_4})
=(\widetilde\Gamma^y_{x_4x_1}\widetilde\Gamma^z_{x_2x_3}+\widetilde\Gamma^y_{x_4x_2}\widetilde\Gamma^z_{x_1x_3}+\widetilde\Gamma^y_{x_4x_3}\widetilde\Gamma^z_{x_2x_1})\widetilde{\mathcal{M}}_{zy}
\end{equation}
However, using the definition of $\widetilde\Gamma = G$, each of the terms on the right-hand side can be manipulated into one of the terms on the left-hand side. For example,
\begin{equation}\label{eq:Gdefn}
\widetilde\Gamma^y_{x_4x_1}\widetilde\Gamma^z_{x_2x_3}\widetilde{\mathcal{M}}_{zy} = G^z_{x_2x_3}(G^y_{x_4x_1}\widetilde{\mathcal{M}}_{zy})= G^z_{x_2x_3}\widetilde{\mathcal{M}}_{zx_1x_4} \equiv \Gamma^z_{x_2x_3}\widetilde{\mathcal{M}}_{zx_1x_4}
\end{equation}
As a result, Eq.~\eqref{eq:GammaM equivalence} is satisfied and we have $\widetilde{\mathcal K}_{x_1x_2x_3x_4} = \widetilde{\mathcal M}_{x_1x_2x_3x_4}$.

 Let us finish the section explaining that even though the introduction of two distinct $(0,2)$ tensors — each inducing its own set of Christoffel symbols — may raise the question of whether there are two `metrics' on the manifold. Yet, one should recall that the definition of covariance on a manifold is independent of any metric structure. Moreover, a connection is fundamentally defined by its transformation behavior, which is constructed to cancel non-tensorial terms. From the $(0,2)$ tensor $T_{ij}$, we define the first connection $\boldsymbol{\Gamma^i_{jk}}$, which is then used to define the second $(0,2)$ tensor $\mathcal N_{ij}$. As we have pointed out,  $\mathcal N_{ij}$ cannot reproduce the correct scattering amplitudes; therefore, a second connection $\Gamma^i_{jk}$ is required, whose components are the Christoffel symbols associated with $\mathcal N_{ij}$. We then construct the modified correlation functions $\mathcal K_{12\cdots n}$ recursively using $\Gamma^i_{jk}$.

\subsection{Arbitrary $n$}\label{sec:induction}
Having worked out the $n=4$ case, let us now generalize our proof of Eq.~\eqref{eq:onshell equiv} to arbitrary $n$. We proceed by induction; with on-shell equivalence established for $n=2, 3, 4$, and assuming that it is valid for $3\leq k\leq n$,
$   \widetilde{\mathcal{K}}_{x_1x_2\cdots x_k}=\widetilde{\mathcal{M}}_{x_1x_2\cdots x_k}$, we want to prove it also holds when $k=n+1$. We will continue to assume that $\widetilde{\mathbf\Gamma}$ is well behaved (has no poles) in the on-shell limit.

We begin by parameterizing $\mathcal{K}_{x_1x_2\cdots x_{n}}$, isolating the pieces containing $\mathcal M_y, \mathcal M_{xy}$: 
\begin{equation}
    \mathcal{K}_{x_1x_2\cdots x_{k}}=\mathcal{M}_{x_1x_2\cdots x_{k}}+U_{x_1x_2\cdots x_{k}}(\mathcal{M}_1)+V_{x_1x_2\cdots x_{k}}(\mathcal{M}_2),
\end{equation}
where $U_{x_1x_2\cdots x_{k}}(\mathcal{M}_1)$ and $V_{x_1x_2\cdots x_{k}}(\mathcal{M}_2)$ are terms containing the 1-point function $\mathcal{M}_y$ and the 2-point function $\mathcal{M}_{xy}$ respectively. By our assumption that on-shell equivalence holds for $k=n$, $\widetilde U = \widetilde V = 0$. We can write the most general $U$ and $V$ terms as~\cite{Cohen:2023ekv}
\begin{equation}\label{eq:parametrization UV}
\begin{split}
    U_{x_1x_2\cdots x_{k}}(\mathcal{M}_1)&=a_{x_1\cdots x_k}^y\mathcal{M}_y+\sum_{i=1}^k b_{x_1\cdots \hat{x_i}\cdots x_k}\mathcal{M}_{x_i},\\
    V_{x_1x_2\cdots x_{k}}(\mathcal{M}_2)&=\sum_{i=1}^kc_{x_1\cdots \hat{x_i}\cdots x_k}^y\mathcal{M}_{\hat{x_i}y}+\sum_{i\neq j}^k d_{x_1\cdots \hat{x_i}\cdots \hat{x_j}\cdots x_k}\mathcal{M}_{x_ix_j},
\end{split}
\end{equation}
where the difference between the $a, b$, etc. terms lies in whether or not the indices on $\mathcal M$ correspond to external legs ($\in \{x_1, \cdots x_k\}$). Inspecting the explicit expressions for $U$ and $V$ when $n=4$ (Eq.~\eqref{eq:Uin4pt}, \eqref{eq:vdefn}), we see that $\mathcal M$ always contains at least one internal index; in other words, $b = d = 0$.

Given $\mathcal K_{x_1x_2\cdots x_n}$, we can calculate the generalized $(n+1)$-point function $\mathcal{K}_{x_1\cdots x_{n+1}}$:
\begin{equation}
\begin{split}
  \mathcal{K}_{x_1\cdots x_{n+1}}  =\, &\, \nabla_{x_{n+1}}\mathcal{K}_{x_1x_2\cdots x_{n}}\\
    =\, &\, \mathcal{K}_{x_1x_2\cdots x_{n},x_{n+1}}-\sum_{i=1}^{n} {\Gamma^y_{x_{n+1}x_i}}\mathcal{K}_{x_1\cdots \hat{x_i}y\cdots x_n}\\
    =\, &\, \frac{\delta}{\delta\phi^{x_{n+1}}}\left[\mathcal{M}_{x_1x_2\cdots x_{n}}+U_{x_1x_2\cdots x_{n}}(\mathcal{M}_1)+V_{x_1x_2\cdots x_{n}}(\mathcal{M}_2)\right]\\
    &\quad-\sum_{i=1}^{n} {\Gamma^y_{x_{n+1}x_i}}\left[\mathcal{M}_{x_1\cdots \hat{x_i}y\cdots x_n}+U_{x_1\cdots \hat{x_i}y\cdots x_n}(\mathcal{M}_1)+V_{x_1\cdots \hat{x_i}y\cdots x_n}(\mathcal{M}_2)\right]\\
    =\,&\,\frac{\delta}{\delta\phi^{x_{n+1}}}\mathcal{M}_{x_1x_2\cdots x_{n}}-\sum_{i=1}^{n} {\Gamma^y_{x_{n+1}x_i}}\mathcal{M}_{x_1\cdots \hat{x_i}y\cdots x_n}+U_{x_1x_2\cdots x_{n+1}}+V_{x_1x_2\cdots x_{n+1}},
\end{split}
\end{equation}
where
\begin{align}
\label{eq:UVdefn}
U_{x_1x_2\cdots x_{n+1}}&=-\sum_{i=1}^{n}\Gamma^y_{x_{n+1}x_i}U_{x_1\cdots \hat{x_i}y\cdots x_n}(\mathcal{M}_1),\\
V_{x_1x_2\cdots x_{n+1}}&=\frac{\delta}{\delta\phi^{x_{n+1}}}\left[U_{x_1x_2\cdots x_{n}}(\mathcal{M}_1)+V_{x_1x_2\cdots x_{n}}(\mathcal{M}_2)\right]-\sum_{i=1}^{n}\Gamma^y_{x_{n+1}x_i}V_{x_1\cdots \hat{x_i}y\cdots x_n}(\mathcal{M}_2).
\end{align}
To prove $\widetilde{\mathcal K}_{x_1 \cdots x_{n+1}} = \widetilde{\mathcal M}_{x_1\cdots x_{n+1}}$, it suffices to show $\widetilde{U}_{x_1x_2\cdots x_{n+1}}=0,\quad \widetilde{V}_{x_1x_2\cdots x_{n+1}}=0$. The first equality is straightforward to see by applying the zero-source condition, while the second is more subtle since it seemingly contains $M_{xyz}$ terms (from the application of the functional derivative to the $M_{x_i y}$ piece of $V$). So, let us examine $V_{x_1x_2 \cdots x_{n+1}}$ more closely, keeping an eye out for terms with multiple internal indices.

All terms in $\frac{\delta U}{\delta\phi^{x_n+1}}$ have at least one external index and therefore vanish on-shell, so the non-trivial terms in $\widetilde V_{x_1x_2\cdots x_{n+1}}$ are:
\begin{align}\label{eq: vanishing V n+1 term}
    \widetilde{V}_{x_1x_2\cdots x_{n+1}}& =\left(\frac{\delta}{\delta\phi^{x_{n+1}}}V_{x_1x_2\cdots x_{n}}(\mathcal{M}_2)-\sum_{i=1}^{n}\widetilde \Gamma^y_{x_{n+1}x_i}V_{x_1\cdots \hat{x_i}y\cdots x_n}(\mathcal{M}_2)\right)
\end{align}
Once we expand the second term using Eq.~\eqref{eq:UVdefn}, we have a double sum:
\begin{align}
\sum\limits_{i=1}^n \Gamma^y_{x_{n+1}x_i}\Big(\sum\limits_{j=1}^n c^z_{x_1\cdots \hat{x}_j \cdots x_n} \widetilde{\mathcal M}_{x_j z}\Big)\Big|_{x_i \to y}.
\end{align}
However, in order for this to survive the on-shell conditions, both $\mathcal M$ indices must be internal and therefore $j = i$. Picking out this piece and combining it with the first term in Eq.~\eqref{eq: vanishing V n+1 term}, we find
\begin{align}
\label{eq:Vtildelast}
    \widetilde{V}_{x_1 \cdots x_{n+1}} & = \sum_{i=1}^nc_{x_1\cdots \hat{x_i}\cdots x_k}^y\widetilde{\mathcal{M}}_{y\,\hat{x_i}x_{n+1}}-\sum_{i=1}^{n}\left(\widetilde \Gamma^z_{x_{n+1}x_i}c_{x_1\cdots \hat{x_i}\cdots x_k}^y\widetilde{\mathcal{M}}_{yz}\right).
\end{align}
Finally, as in Eq.~\eqref{eq:Gdefn}, we can set $\widetilde \Gamma \equiv G$ and use the definition of $G$ to convert
\begin{align}
\widetilde \Gamma^y_{x_{n+1}x_i} \widetilde{\mathcal{M}}_{yz} \equiv G^y_{x_{n+1}x_i} \widetilde{\mathcal{M}}_{yz} = \widetilde{\mathcal M}_{x_{n+1}x_i\, z}.
\end{align}
As $\mathcal M_{xyz}$ is symmetric under interchanging indices by definition, Eq.~\eqref{eq:Vtildelast} vanishes and we have $\widetilde{\mathcal{K}}_{x_1x_2\cdots x_{n+1}}=\widetilde{\mathcal{M}}_{x_1x_2\cdots x_{n+1}}$ for arbitrary $n \ge 4$.

\section{Conditions for well behaved $\widetilde{\mathbf \Gamma}$}\label{sec: necessary conditions}

Our results to this point have required both the existence of a $(0,2)$ tensor $T$ {\it and} that the connection derived from this tensor is well behaved in the on-shell limit. We showed that the first requirement is easily satisfied, as the kinematic metric  $g_{ij}$ transforms as required. In this section, we explore the second criteria in more detail, and determine what limitations well-behaved $\widetilde{\mathbf \Gamma}$ place on theories.

To make things more concrete, we work with an example -- a massive real scalar field theory with a non-derivative trilinear,
\begin{equation}
    \mathcal L=\frac{1}{2}\partial\phi\partial\phi-\frac{1}{2}m^2\phi^2-\frac{1}{6}\lambda\phi^3.
\end{equation}
This is the nearly the same setup as considered in Appendix A of Ref.~\cite{Cohen:2024bml}, which will allow us to borrow many of their results. More detailed steps of the following derivations can also be found in Appendix ~\ref{sec: massive}. To facilitate the connection with Ref.~\cite{Cohen:2024bml,Cohen:2025prs}, we will work in momentum space. We will use the same symbols (e.g. $\mathcal M$ or $\mathbf{\Gamma}$), but these should be understood now as momentum-dependent.

Following the techniques of Ref.~\cite{Cohen:2024bml}, we can extract the (momentum space) $(0,2)$ tensor $T = g_{ij}$ and the tensor $\mathbf \Gamma$ formed from it. Evaluated at the vacuum of the theory $\phi = 0$, we find
\begin{equation}
    \mathbf{\Gamma^w_{xy}}=-(2\pi)^4\delta(p_w-p_x-p_y)\frac{p_w^2}{p_w^2-m^2}\frac{1}{6}\lambda\left(\frac{1}{p_wp_x}+\frac{1}{p_xp_y}+\frac{1}{p_yp_w}\right).
\end{equation}
The combination appearing in higher point amplitudes, such as Eq.~\eqref{eq:expandK}, is \begin{equation}\label{eq: necessary condition}
    \boldsymbol{\Gamma}^a_{bc}\mathcal M_{ad},
\end{equation}
where $\mathcal M_{ad}$ is the tree-level two point amplitude (also evaluated at $\phi = 0$)
\begin{equation}
    \mathcal M_{ad}=(2\pi)^4\delta(p_a+p_d)(p_a^2-m^2).
\end{equation}
Forming $\mathbf{\Gamma^a_{bc}} \mathcal M_{ad}$, we see the $(p^2_a - m^2)$ term in $\mathcal M$ -- which gets set to zero when we take all external momenta on shell -- cancels with the denominator of $\mathbf{\Gamma}$. As a result, $\mathbf{\Gamma^a_{bc}} \mathcal M_{ad} \propto \lambda\, p^2_a \sim \lambda m^2$ after using $p_a = -p_d$ and setting the external momentum ($p_d$) on-shell. 

With $\boldsymbol{\Gamma}^a_{bc}\mathcal M_{ad}|_{p_d^2=m^2} \ne 0$, in order  to recover the results of Sec.~\ref{sec:onshell} and \ref{sec:induction}, we must restrict ourselves to setups where $\lambda\, m^2 = 0$ -- but does this mean the mass must be zero, the triliear coupling, or both? 

If the scalar is massless but $\lambda \ne 0$, it would seem as if the issue is resolved, as 
\begin{equation}\label{eq: appendix wrong zero condition}
    \boldsymbol{\Gamma}^w_{xy}\mathcal M_{wz}\sim p_z^2\lambda\left(\frac{1}{p_zp_x}-\frac{1}{p_xp_y}+\frac{1}{p_yp_z}\right)\overset{p_z^2=0}{\longrightarrow}0.
\end{equation}
However, once we sum over the full set of $\mathbf\Gamma \mathcal M$ terms that appear in the three-point vertex (Eq.~\eqref{eq:Nabc}) and impose momentum conservation $p_x+p_y+p_z=0$, we find
\begin{align}
    \boldsymbol{\Gamma}^w_{xy}\mathcal M_{wz}+\boldsymbol{\Gamma}^w_{zx}\mathcal M_{wy}+\boldsymbol{\Gamma}^w_{yz}\mathcal M_{wx}
    & \propto \lambda\left[p_x^2\left(\frac{1}{p_zp_x}+\frac{1}{p_xp_y}-\frac{1}{p_yp_z}\right)+perm(x,y,z)\right]\\
    & \propto \lambda\left[\frac{p_x^2+p_y^2-p_z^2}{p_xp_y}+perm(x,y,z)\right]\\
    & \propto 6\,\lambda,
\end{align}
another non-zero result. However, we have been too hasty. We have assumed throughout this section that $\phi = 0$ is the true (tree-level) minimum of the theory -- but this is not the case for a massless theory with a $\phi^3$ interaction, which only has a saddle point at $\phi = 0$. If we try to fortify the theory by adding higher order polynomial terms, that fixes the runaway behavior but also inevitably leads to a lower minimum at $\phi^* \ne 0$.\footnote{As an example, consider $\mathcal L=\frac{1}{2}\partial\phi\partial\phi-\frac{1}{6}\lambda\phi^3-\frac{1}{24}g\phi^4,\ g>0$. We can easily verify that $\phi=0$ is the ``false vacuum" since the second derivative at this point vanishes, and the ``true vacuum" locates at $\phi^*=-\frac{3\lambda}{g}$ with a physical mass $m^*=\frac{3\lambda^2}{2g}$.} Excitations about $\phi^*$ are massive. Therefore, in order to consider a consistent massless theory, we must set $\lambda = 0$.

With $\lambda = 0$, what interactions can there be? Higher point (quartic, etc.) interactions pose no threat. Following the steps of Sec.~\ref{sec: covariant functional manifolds}, these interactions can be combined into $g_{ij}$ and will appear in $\mathbf{\Gamma}$. At the vacuum field value, these will vanish as they have too many fields to contribute to the three-point vertex. Higher ($n>4$) point amplitudes involve the functional derivatives of $\mathbf{\Gamma}$, which are non-zero at the vacuum. These contain denominator structures which can cancel the behavior of $\mathcal M_{xy}$ on-shell. However, the contribution of higher point interactions to  $\mathbf \Gamma$ (and it's functional derivatives) are always $\propto p^2_z$ where $p_z$ is an external (massless) momenta. As there are no special kinematics for $n>3$ particle vertices, the logic of Eq.~\eqref{eq: appendix wrong zero condition} holds.~\footnote{By the same reasoning, we can see that our formalism also fails for massive scalar theories with interactions $\lambda \phi^n, n>3$. For the massive case, we again find the on-shell limit of the (derivatives of the) connection contracted with $\mathcal M_{xy}\, \propto\, \lambda p^2_z$ but since the fields are massive, this is $\lambda m^2$. The source of the issue is not the interaction, but the inverse metric, which is singular for massive fields in the on-shell limit (see Eq. (A.11) of Ref.~\cite{Cohen:2024bml}) .} This also matches Eq. (24) of \cite{Cohen:2024bml}. Derivative, trilinear interactions, of the form $\phi (\partial_\mu \phi )(\partial^\mu \phi)$ are also allowed. 

From a geometric point of view, issues with the $\phi^3$ interaction can be traced back to the fact that there does not exist any local field redefinition to remove it if $m=0$, i.e. evaluate in the Riemann normal coordinates such that the connection vanishes.\footnote{We should emphasize that this caveat can be removed if we work with a connection without the mass pole, and such possibilities are left for future studies.} A non-local field redefinition can make it possible, but it is incompatible with the LSZ reduction formula and changes the physical on-shell S-matrix.


\section{Geometric interpretation}\label{sec: geometric interpretation}

 As we have shown, a massless theory without $\phi^3$ interaction is necessary to establish a covariant off-shell recursion relation. Within this restricted class of theories, we introduce the connection $\Gamma^z_{xy}$ and modified correlation functions $\mathcal K$, such that $\mathcal K_{n+1}=\nabla_{n+1}\mathcal{K}_n$. The corresponding Riemann curvature is straightforward to calculate:
\begin{equation}
    R_{abcd}=\frac{1}{2}(-\mathcal{N}_{bd,ca}+\mathcal{N}_{bc,da}-\mathcal{N}_{ac,db}+\mathcal{N}_{ad,cb})+\mathcal{N}_{ef}(\Gamma^e_{da}\Gamma^f_{cb}-\Gamma^e_{ca}\Gamma^f_{db}).
\end{equation}
Since our goal is to relate the curvature tensor to scattering amplitudes, let us impose on-shell conditions on both sides
\begin{equation}
\begin{split}
    \widetilde{R}_{abcd}&=\frac{1}{2}(G^e_{bd}\widetilde{\mathcal{M}}_{eca}-G^e_{bc}\widetilde{\mathcal{M}}_{eda}+G^e_{ac}\widetilde{\mathcal{M}}_{edb}-G^e_{ad}\widetilde{\mathcal{M}}_{ecb})\\&\ \ \ +\widetilde{\mathcal{M}}_{ef}(G^e_{da}G^f_{cb}-G^e_{ca}G^f_{db})\\
    &=0
    \end{split}
\end{equation}
where we note that
\begin{equation}
   \widetilde{\mathcal{N}}_{bd,ca}=\widetilde{\mathcal{M}}_{bdca}-G^e_{bd}\widetilde{\mathcal{M}}_{eca},
\end{equation}
and Eq. \eqref{eq:Gdefn} is used.

Consequently, we conclude that the (on-shell) curvature induced by $\mathcal{K}$ is zero. In fact, in \cite{Cohen:2023ekv} the authors show that the on-shell curvature induced by $\mathcal M$ is zero, using the crossing symmetry between the last two legs $x_{n+1}$ and $x_{n+2}$:
\begin{equation}\label{eq: zero curvature}
    \mathcal{M}_{x_1\cdots x_nx_{n+1}x_{n+2}}= \mathcal{M}_{x_1\cdots x_nx_{n+2}x_{n+1}}\Rightarrow[\nabla_{x_{n+1}},\nabla_{x_{n+2}}] \mathcal{M}_{x_1\cdots x_n}=0.
\end{equation}
We then arrive at the same conclusion by noting that $\widetilde {\mathcal K}=\widetilde{\mathcal M}$. It is worth noting that the functional geometry may exhibit nonzero curvature away from the physical point. However, when expressing scattering amplitudes geometrically, only the geometry at the physical point is relevant. That being said, in the rest of this section, we refer to this local flatness when we use the word ``flat''. 

The vanishing curvature seems to strongly disfavor a geometric description of the functional manifold because it is expected that when we ignore higher-derivative interactions, the functional approach reduces to the field-space method, whose curvature is non-trivial and has important physical implications, serving as a key ingredient in expressing $4$-point amplitudes and beyond. 

At first glance, this seems to be a contradiction, but it can be resolved based on the observation that an m-sphere $S^m$ which is embedded in a flat Euclidean space $R^{n>m}$\footnote{Note that there is a subtle difference here: the functional manifold is infinite dimensional while the dimension of $R^{n}$ is still finite. In addition, it is not clear how to define a reasonable curvature on an infinite-dimensional manifold. Nevertheless, we shall simply use this example as an analogy and ignore the technical details.}, is equipped with a nonzero curvature. The embedding is realized by freezing certain coordinates; for example, one fixes the radius $r=R$ when $n=m+1$. The same trick also works in our case. Specifically, the functional manifold is spanned by $\{\phi,\partial\phi,\partial^2\phi,\cdots\}$, while the field space is spanned by $\{\phi\}$. By freezing the derivative coordinates, the flat functional manifold reduces to some lower-dimensional submanifold with non-vanishing curvature, which in our example going from the functional-space (flat) to the field-space. In this sense, we claim that the functional method reduces to the usual field-space if we focus on non-derivative field redefinitions (setting $\partial\phi=0$) and impose on-shell conditions from a geometric point of view. Indeed, as has been shown in Ref. \cite{Cohen:2023ekv}, the ``on-shell connection'' $G^c_{ab}$ of the functional manifold ultimately reduces to the Christoffel symbols $\gamma^c_{ab}$ of the field-space geometry once we consider the constant field configuration, i.e. $\partial\phi=0$, which can be expressed in terms of the sequence $\Gamma^c_{ab}\xrightarrow{on-shell}G^c_{ab}\xrightarrow{\partial\phi=0}\gamma^{c}_{ab}$. We emphasize that this does not mean that the two manifolds (functional geometry, on shell vs. field space) are equivalent, just that the two connections are equal as functions.
We generalize the geometry of field-space to the functional manifold and perform calculations within this broader structure. This is analogous to working with the flat space $R^n$ rather than the constrained sphere $S^m$. \footnote{The recent paper \cite{Cohen:2025prs} also finds a particular choice of metric that relates functional geometry and field-space geometry, from which one can reconstruct on-shell covariant vertex functions.} 

In addition, this perspective suggests a deeper relation between the infinite-dimensional functional manifold and jet bundle geometry. For example, if we only require $\partial^2\phi=0$ and leave the first derivative coordinate $\partial\phi$ free in the manifold, we expect that the resulting geometry is related to the Lagrange space or 1-jet bundle. We anticipate this pattern to hold up to the $\infty$-jet as well and thus allows for a more rigorous construction using the jet bundle language. However, this topic is beyond the scope of this paper and is a project in progress for future publication.

Having  provided some evidence that curved field-space geometry may be embedded in a locally flat functional manifold, a natural question arises: in the absence of Riemann curvature, what serves as the fundamental object for scattering amplitudes? Conventionally, both the field-space and kinematics-geometry duality rely on the metric and its curvature as the foundational building blocks. We argue that this reliance is unnecessary once we loose the meaning of geometry a little bit. The key is not the existence of a nonzero curvature, but a consistent covariant framework with tensorial properties. By slightly broadening the meaning of ``geometry," we demonstrate that a well-chosen metric and connection alone are sufficient to construct the amplitudes, rendering a curved background superfluous for this task. 

To illustrate this point, we summarize and compare the essential building blocks of both the functional approach and the field-space approach in Table \ref{tab:comparison}. The derivatives are defined by the underlying manifold, the functional derivative and partial derivative respectively. The metrics, on the other hand, are not unique and are defined from  scalar quantities, which we choose to be the effective action on the functional manifold and the Lagrangian on the field-space. Both field and functional geometries employ the torsion-free Levi-Civita connection, which is the standard choice for Riemannian geometry. The 4-point amplitudes on the functional manifold is determined by the modified $4$-pt correlation function $\mathcal K_{abcd}$, while in the case of field-space is a combination of curvature and Mandelstam variables.

Our perspective is that the primary focus should be on the connection and the associated covariant derivative rather than on the curvature tensor. The key insight is that a fully covariant framework for amplitudes can be built recursively by acting with covariant derivatives. The recursion relation is naturally encoded in our approach, and it yields modified $n$-pt correlation functions, making the presence of a vanishing curvature a secondary concern. This method of geometric realization is innovative and differs from the construction in field-space. This perspective is similar to what the authors in Ref.~\cite{Cohen:2025prs} advocate, and interested readers may refer to the paper for details.

Before we end, we want to point out that having an off-shell covariant formulation may provide a better understanding of the structure of EFTs. For example, there are terms like $\nabla_e R_{abcd},\ \nabla_f\nabla_e R_{abcd},\ \cdots$ that are not necessarily trivial, even at the vacuum point, and may have physical implications independent of the fact that the local curvature tensor vanishes. An off-shell formalism is preferred over the on-shell construction along this line because it keeps everything manifestly covariant. A similar story occurs in the Lagrange space approach, where the authors noticed that even though the (hv-)torsion $\mathcal P^\alpha_{\beta\gamma}$ vanishes at the vacuum, its (h-)covariant derivative $\mathcal P^\alpha_{\beta\gamma/\delta}$ is related to the four-point Wilson coefficient. Ref. \cite{Craig:2023wni} provides a rigorous derivation and precise definitions for interested readers. Finally,  when adding fermions and/or gauge bosons to the picture,  our approach  will not need any new `anholonomic' terms. 
\begin{table}[h]
    \centering
    \begin{tabular}{|c|c|c|}
        \hline
         & Functional Manifold & Field-space Geometry \\ \hline
        Vector/Derivative & $\frac{\delta}{\delta\phi}$ & $\frac{\partial}{\partial\phi}$ \\ \hline
        Scalar & $\Gamma[\phi]$ (Eq.~\ref{eq: effective action})& $\mathcal{L}(\phi,\partial\phi)$ \\ \hline
        Metric & $\mathcal{N}_{ab}$ (Eq.~\ref{eq:Nabdefn}) & $g_{ab}$ \\ \hline
        Connection & $\Gamma^c_{ab}$ (Eq.~\ref{eq:Gammadef}) & $\gamma^c_{ab}$ \\ \hline
        Covariant derivative & $\nabla\sim\nabla(\mathcal{N}_{ab},\Gamma^c_{ab})$ & $\nabla\sim\nabla(g_{ab},\gamma^c_{ab})$ \\ \hline
        4-point amplitude & $\mathcal{A}_{abcd}\sim\mathcal{K}_{abcd}\sim\nabla_d\nabla_c\mathcal{K}_{ab}$ & $\mathcal{A}_{abcd}\sim {R}_{abcd}u+{R}_{acbd}s$ \\ \hline
    \end{tabular}
    \caption{Comparison of quantities in field-space vs. functional geometry}
    \label{tab:comparison}
\end{table}
 
\section{Extensions and future directions}\label{sec: future}

 In this paper, we have developed a manifest off-shell covariant geometric description of scattering amplitudes on the functional manifold for a general massless scalar theory at tree-level\footnote{Recall that, in order to be consistent, a massless scalar theory cannot have a $\phi^3$ term.}. By introducing a ``good'' connection and the associated covariant derivative, the `anholonomic' terms in the on-shell recursion relation are removed. 

 To achieve off-shell covariance, we `improve' the correlation functions using $\mathbf{\Gamma}$ the Christoffel symbols formed from a genuine $(0,2)$ tensor on the manifold. A natural $(0,2)$ tensor, and the one we utilize, is $g_{ij}$, the prefactor of the kinetic term.  The improved correlation functions $\mathcal N$ are covariant, but do not satisfy the expected recursive relation linking smaller correlation functions to larger ones. We showed that the desired recursive properties can be achieved if we further modify the correlation functions $\mathcal N \to \mathcal K$, treating the two point $\mathcal N$ as a metric and lumping the functional derivative with the connection formed from $\mathcal N$ into a covariant derivative. The alterations to the correlation functions all vanish when evaluated at the vacuum and after taking the on-shell limit, so $\widetilde {\mathcal K} = \widetilde{\mathcal M}$ provided the Christoffel $\mathbf{\Gamma}$ are not singular in that limit. This requirement restricts the formalism to massless scalar theories.
 
 For this restricted set of theories, we then postulate further relations among the functional manifold, the existing field-space geometry, the Lagrange space, and the jet-bundle formalism. A local ``flat'' zero curvature manifold is acceptable in the sense that it is understood as an infinite-dimensional extension of a curved field-space geometry. This is in line with our perspective laid out in Section \ref{intro}, where we emphasize that general covariance under field redefinitions is sufficient for a geometric description, whether or not the underlying manifold is curved, i.e. equipped with a nonzero curvature.

From a mathematical perspective, the functional geometry approach only requires the following building blocks: coordinates $[\phi^x]$, functional derivatives $\frac{\delta}{\delta\phi}$, and a scalar function $\Gamma[\phi]$ depending on the entire field configurations. This way of understanding geometric amplitudes is innovative, since it is independent of an initial metric. Recall that one of the difficulties in extending the field-space geometry to include higher-derivative interactions is the construction of a metric. In this paper, the metric and the associated covariant derivative are induced from the effective action alone -- rather than a priori two-derivative Lagrangian -- at the cost of introducing additional structures: $T_{xy},\mathcal N_{xy}$ and $\mathcal K_{xy}$. Let us reemphasize that the fact that we have introduced two different $(0,2)$ tensors does not imply any problem in our geometric approach, since we only require amplitudes to transform covariantly under a field redefinition.

We expected that massless fermions and gauge bosons could be incorporated into our formalism following a similar construction to what we have shown, and that the resulting geometry should reduce to the well-established field-space framework when constant field configuration is imposed. In addition, since the only requirement of the ``scalar'' term is how it transforms under general field redefinitions, we can replace the tree-level action with the 1-loop effective action $\Gamma_{1-loop}$
\begin{equation}
    \Gamma_{1-loop}[\phi]=S[\phi]+\frac{i}{2}\log\text{Det}\left(\frac{\delta^2 S[\phi]}{\delta\phi^{x_1}\delta\phi^{x_2}}\right),
\end{equation}
and the corresponding geometry shall describe the scattering amplitudes at 1-loop. On the other hand, we expect that significant modifications will be needed in order to extend the present formalisms to massive theories.

\appendix 

\section*{Acknowledgments}

We thank Beno\^{\i}t Assi, Tim Cohen and Kevin Zhang for carefully reading the manuscript and giving useful comments. We are especially grateful to Kevin Zhang for bringing the issue of poles in $\mathbf{\Gamma}$ to our attention. This work is partially supported by the National Science Foundation under grant PHY-2412701. R.W is also supported by the National Science Foundation of China under Grant No. 12547137 and the China Postdoctoral Science Foundation under Grant Number 2025M783371.

\section{Zero source and on-shell conditions}\label{sec: on-shell conditions}
To go on-shell, we Fourier transform the correlation functions after evaluating them in the vacuum of the theory, then perform the LSZ reduction. As reviewed in Ref.~\cite{Cohen:2023ekv}, we consider a scalar theory with action $S[\phi]$, which can be fully encoded into a partition function $Z[J]$ using the path-integral formalism:
\begin{equation}
    Z[J]=\int \mathcal{D}\phi e^{iS[\phi]+i\int d^4x \phi^xJ^x},
\end{equation}
where $J^x$ is a classical external current. $Z[J]$ is the generating functional of the $J$-dependent correlation function:
\begin{equation}
    \langle\phi^{x_1}\phi^{x_2}\cdots\phi^{x_n}\rangle_J\equiv (-i)^n\frac{1}{Z[J]}\frac{\delta^n Z[J]}{\delta J^{x_1}\delta J^{x_2}\cdots\delta J^{x_n}}.
\end{equation}
Another useful quantity $W[J]$, the generating functional for connected correlation functions:
\begin{equation}
    \langle\phi^{x_1}\phi^{x_2}\cdots\phi^{x_n}\rangle_{J,connected}\equiv (-i)^{n+1}\frac{\delta^nW[J]}{\delta J^{x_1}\delta J^{x_2}\cdots\delta J^{x_n}}.
\end{equation}
The one-particle-irreducible (1PI) effective action $\Gamma[\phi]$ is defined as a Legendre transformation of $W[J]$:
\begin{equation}\label{eq: effective action}
    \Gamma[\phi]\equiv W[J[\phi]]-\int d^4x J[\phi]\phi(x).
\end{equation}
Taking the functional derivative of $\Gamma[\phi]$ yields the source term $J[x]$:
\begin{equation}\label{eq:sourceterm}
    J_x=\frac{\delta \Gamma[\phi]}{\delta \phi^x},
\end{equation}
and the zero source condition is correspondingly defined as:
\begin{equation}\label{eq:zerosource}
    \widetilde{\mathcal M}_x\equiv\frac{\delta \Gamma[\phi]}{\delta \phi^x}\bigg|_{\phi=\phi_v}=0,
\end{equation}
where $\phi_v(x)$ is the quantum VEV of the scalar fields for the original theory.

The second functional derivative of $\Gamma[\phi]$ defines the propagators $D(x,y)$:
\begin{equation}\label{eq:propa}
    D^{-1}(x,y)=-\left[\frac{\delta^2 W[J]}{\delta J^x\delta J^y}\right]^{-1}=-\frac{\delta^2\Gamma[\phi]}{\delta \phi^x\delta\phi^y}.
\end{equation}
The on-shell condition is easily imposed when we move to momentum space:
\begin{equation}\label{eq:onshellcondition}
    \widetilde{\mathcal M}_{xy}\equiv\int d^4 x e^{i\bar{p} x}\frac{\delta^2\Gamma[\phi]}{\delta \phi^x\delta\phi^y}\bigg|_{\phi=\phi_v}=0,
\end{equation}
where $\bar{p}^2\equiv m^2=0$ is the on-shell momentum of massless scalar fields. Since our main interest in this paper is the tree-level amplitude, $\Gamma[\phi]=S[\phi]=\int d^4x \mathcal{L}[\phi(x)]$.

\section{Example}\label{sec: appendix}
In this section, we apply the formalism constructed in this paper to a concrete example to show that $\widetilde {\mathcal K}_{ab}=\widetilde {\mathcal M}_{ab}$. The Lagrangian of the theory is given by:
\begin{equation}
    \mathcal{L}=\frac{1}{2}(1+\epsilon\phi)\partial\phi\partial\phi.
\end{equation}
The corresponding $\mathcal{M}_x$ and $\mathcal{M}_{xy}$ are:
\begin{equation}
\begin{split}
    \mathcal{M}_x&=-\left(1 + \epsilon \phi(x)\right) \Box \phi(x) - \frac{\epsilon}{2} \partial \phi(x) \partial \phi(x),\\
    \mathcal{M}_{xy}&=-\epsilon \delta(x - y) \Box \phi(x)- \left(1 + \epsilon \phi(x)\right) \Box \delta(x - y) - \epsilon \partial \phi(x) \partial \delta(x - y),
\end{split}
\end{equation}
where the subscripts $x,xy$ refer to the position in which the function/functional is evaluated and $\Box\equiv \partial^2$ is the d'Alembertian. The $(0,2)$ tensor $T_{xy}$ for this theory is extracted from the Lagrangian:
\begin{equation}
    T_{xy}=-\left(1 + \epsilon \phi(x)\right) \delta(x - y),
\end{equation}
and its inverse $T^{xy}$:
\begin{equation}
    T^{xy}=-\frac{1}{1 + \epsilon \phi(x)} \delta(x - y), 
\end{equation}
from which we can calculate $\mathbf{\Gamma^x_{yz}}$:
\begin{equation}
\begin{split}
    \mathbf{\Gamma^x_{yz}}&=\frac{1}{2}T^{xw}(T_{wy,z}+T_{wz,y}-T_{yz,w})\\
    &=\frac{1}{2}\int dw\frac{\epsilon}{1 + \epsilon \phi(x)} \delta(x - w)[2\delta(w-z)\delta(w-y)-\delta(y-w)\delta(y-z)]\\      
    &=\frac{\epsilon}{2 \left[1 + \epsilon \phi(x)\right]} \delta(x - y) \delta(x - z)
\end{split}
\end{equation}
Note that we use $dw\equiv d^4w$ and $\delta(x-y)\equiv \delta^4(x-y)$ for simplicity.

It is straightforward to calculate $\mathcal{N}_{xy}$:
\begin{equation}
    \begin{split}
        \mathcal{N}_{xy} &=\mathcal{M}_{xy}-\mathbf{\Gamma^z_{xy}}\mathcal{M}_z\\
        &= -\frac{\epsilon}{2} \delta(x - y) \Box \phi(x)- \left[1 + \epsilon \phi(x)\right] \Box \delta(x - y)- \epsilon \partial \phi(x) \partial \delta(x - y) \\
\quad &+ \frac{\epsilon^2}{4 \left[1 + \epsilon \phi(x)\right]} \delta(x - y) \partial \phi(x) \partial \phi(x)\\
&=-\Box \delta(x - y)-\epsilon\left[\frac{1}{2} \delta(x - y) \Box \phi(x)+\phi(x)\Box \delta(x - y)+\partial \phi(x) \partial \delta(x - y)\right]\\
&+\frac{\epsilon^2}{4 \left[1 + \epsilon \phi(x)\right]} \delta(x - y) \partial \phi(x) \partial \phi(x)\\
    &\equiv \mathcal{N}^{(0)}_{xy}+\epsilon\mathcal{N}^{(1)}_{xy}+\epsilon^2\mathcal{N}^{(2)}_{xy},
\end{split}
\end{equation}
where we group the terms by order in $\epsilon$. It is easy to see that $\overline{\mathcal{N}}_{xy}=\overline{\mathcal{M}}_{xy}$ once we set $\mathcal{M}_x=0$.

We will derive the inverse $\mathcal{N}^{xy}$ perturbatively in $\epsilon$, i.e. let $\mathcal{N}^{xy}=\mathcal{N}^{(0)xy}+\epsilon\mathcal{N}^{(1)xy}+\mathcal{O}(\epsilon^2)$, such that
\begin{equation}
    \int dy\mathcal{N}_{xy}\mathcal{N}^{yz}=\delta^z_x.
\end{equation}
We only need the first two pieces for future purposes:
\begin{equation}
\begin{split}
    \mathcal{N}^{(0)xy}&=G(x-y),\\
    \mathcal{N}^{(1)xy}&=-\phi(x)G(x-y)+\frac{1}{2}\int dz G(x-z)\left[\Box \phi(z)G(z-y)+2\partial\phi(z)\partial G(z-y)\right],
\end{split}
\end{equation}
where $G(x-y)$ is the Green's function satisfying $-\Box G(x-y)=\delta(x-y)$.
Note that the inverse of $\mathcal{M}_{xy}$ can be derived in a similar manner:
\begin{equation}
\begin{split}
\mathcal{M}^{xy}&=\mathcal{M}^{(0)xy}+\epsilon\mathcal{M}^{(1)xy}+\mathcal{O}(\epsilon^2),\\
\mathcal{M}^{(0)xy}&=G(x-y),\\
\mathcal{M}^{(1)xy}&=-\phi(x)G(x-y)+\int dz G(x-z)\left[\Box \phi(z)G(z-y)+\partial\phi(z)\partial G(z-y)\right].
\end{split}
\end{equation}
If we impose $\mathcal{M}_x=0$ up to $\mathcal{O}(\epsilon^0)$, we get $\Box\phi=0$ and $\overline{\mathcal{N}}^{xy}\overset{\mathcal{O}(\epsilon)}{=}\overline{\mathcal{M}}^{xy}$. This validates our earlier derivation in Eq. \eqref{eq:onshellconnect2}.

Now let us derive the Christoffel symbol $\Gamma^x_{yz}$
\begin{equation}
\begin{split}
    \Gamma^x_{yz}&=\frac{1}{2}\mathcal{N}^{xw}(\mathcal{N}_{wy,z}+\mathcal{N}_{wz,y}-\mathcal{N}_{yz,w})\\
    &=\epsilon\Gamma^{(1)x}_{yz}+\mathcal{O}(\epsilon^2),
\end{split}
\end{equation}
where
\begin{equation}
\begin{split}
    \Gamma^{(1)x}_{yz}&=\frac{1}{2}\mathcal{N}^{(0)xw}(\mathcal{N}^{(1)}_{wy,z}+\mathcal{N}^{(1)}_{wz,y}-\mathcal{N}^{(1)}_{yz,w})\\
    &=-\frac{1}{2}\int dwG(x-w)\left[\frac{1}{2} \delta(w-y) \Box \delta(w-z)+\delta(w-z)\Box \delta(w-y)+\partial \delta(w-z) \partial \delta(w-y)\right.\\
&\ \left. + \frac{1}{2} \delta(w-z) \Box \delta(w-y)+\delta(w-y)\Box \delta(w-z)+\partial \delta(w-y) \partial \delta(w-z)\right.\\
&\ \left. -\frac{1}{2} \delta(y-z) \Box \delta(y-w)-\delta(y-w)\Box \delta(y-z)-\partial \delta(y-w) \partial \delta(y-z)\right]\\
&=-\frac{1}{2}\int dwG(x-w)\bigg[3\delta(w-y)\Box\delta(w-z)+2\partial \delta(w-z) \partial \delta(w-y)\\
&-\frac{3}{2}\delta(y-z)\Box\delta(y-w)-\partial\delta(y-w)\partial\delta(y-z)\bigg].
\end{split}
\end{equation}

Finally, we can calculate $\mathcal{K}_{xy}=\mathcal{M}_{xy}-\Gamma^z_{xy}\mathcal{M}_z$ at order $\mathcal{O}(\epsilon)$:
\begin{equation}
    \begin{split}
        \mathcal{K}_{xy}&\overset{\mathcal{O}(\epsilon)}{=}-\epsilon \delta(x - y) \Box \phi(x)- \left(1 + \epsilon \phi(x)\right) \Box \delta(x - y) - \epsilon \partial \phi(x) \partial \delta(x - y)\\
        &-\frac{1}{2}\int dwG(z-w)\bigg[3\delta(w-x)\Box\delta(w-y)+2\partial \delta(w-y) \partial \delta(w-x)\\
&-\frac{3}{2}\delta(x-y)\Box\delta(x-w)-\partial\delta(x-w)\partial\delta(x-y)\bigg][\left(1 + \epsilon \phi(z)\right) \Box \phi(z) + \frac{\epsilon}{2} \partial \phi(z) \partial \phi(z)].
    \end{split}
\end{equation}

It is unnecessary to evaluate the bulky integral since setting $\mathcal{M}_z=0$ will kill the entire term, and we find $\overline{\mathcal{K}}_{xy}=\overline{\mathcal{M}}_{xy}$ up to $\mathcal{O}(\epsilon)$. The covariant property of the higher-point amplitudes $\mathcal{K}_{n}$ is ensured by construction.\footnote{The proof is given in the main text, and we shall not repeat it here.}

\section{Singularity in massive theories}\label{sec: massive}

In this appendix, we give a more thorough investigation of the singularities in massive theories, and show that we need to impose $\lambda m^2=0$ for Eq. \eqref{eq:onshellconnect2} to hold. 
Note that we choose to work in momentum space to be in line with the results in \cite{Cohen:2024bml}. We will still use the same symbols, but it is understood that they are now momentum-dependent after the Fourier transform.

We assume a massive scalar theory
\begin{equation}
    \mathcal L=\frac{1}{2}\partial\phi\partial\phi-\frac{1}{2}m^2\phi^2-\frac{1}{6}\lambda\phi^3,
\end{equation}
from which $T_{xy}$ can be read off:
\begin{equation}\label{eq:appendix T}
    T_{xy}=(2\pi)^4\delta(p_x+p_y)\left(1+\frac{m^2}{p_xp_y}\right)+\frac{1}{3}\lambda\frac{1}{p_xp_y}\phi(-p_x-p_y).
\end{equation}
The inverse $T^{xy}$ evaluated at the vacuum is
\begin{equation}
    T^{xy}=(2\pi)^4\delta(p_x+p_y)\frac{p_x^2}{p_x^2-m^2}.
\end{equation}
Now let us compute $\mathbf{\Gamma^w_{xy}}$, $\mathcal M_{xy}$ and $\mathcal M_{xyz}$ at $\phi=0$:
\begin{equation}\label{eq: appendix Gamma}
    \mathbf{\Gamma^w_{xy}}=-(2\pi)^4\delta(p_w-p_x-p_y)\frac{p_w^2}{p_w^2-m^2}\frac{1}{6}\lambda\left(\frac{1}{p_wp_x}+\frac{1}{p_xp_y}+\frac{1}{p_yp_w}\right),
\end{equation}
\begin{equation}\label{eq: appendix M2}
    \mathcal M_{xy}=(2\pi)^4\delta(p_x+p_y)(p_x^2-m^2),
\end{equation}
\begin{equation}\label{eq: massive 3-pt}
    \mathcal M_{xyz}=-(2\pi)^4\delta(p_x+p_y+p_z)\lambda.
\end{equation}

We are now able to compute the on-shell limit (taking $x_1$ and $x_2$ to be on-shell) of Eq. \eqref{eq:Gammadef},
\begin{equation}\label{eq: massive on-shell gamma}
    \widetilde \Gamma_{x_1 x_2}^y = \frac{1}{2}\left(\widetilde {\mathcal{M}}^{-1}\right)^{y z}\left(\widetilde {\mathcal{N}}_{x_1 z , x_2}+\widetilde {\mathcal{N}}_{x_2 z , x_1}-\widetilde {\mathcal{N}}_{x_1 x_2 , z}\right)
\end{equation}
Let us compute the first term :
\begin{equation}
\begin{split}
    \widetilde {\mathcal{N}}_{x_1 z , x_2}&=\widetilde {\mathcal{M}}_{x_1zx_2}-\widetilde{\mathbf{\Gamma}}^w_{x_1z}\widetilde{\mathcal M}_{wx_2}\\
    &=\widetilde {\mathcal{M}}_{x_1zx_2}+(2\pi)^4\frac{\lambda p_{x_2}^2}{6}\left(\frac{1}{-p_{x_2}p_{x_1}}+\frac{1}{-p_{x_2}p_{z}}+\frac{1}{p_zp_{x_1}}\right)|_{p_{x_1}^2=p_{x_2}^2=m^2}\\
    &=\widetilde {\mathcal{M}}_{x_1zx_2}-(2\pi)^4\frac{\lambda m^2}{6}\left(\frac{1}{p_{x_2}p_{x_1}}-\frac{1}{p_{x_2}(p_{x_2}+p_{x_1})}+\frac{1}{p_{x_1}(p_{x_2}+p_{x_1})}\right)|_{p_{x_1}^2=p_{x_2}^2=m^2}\\
    &=\widetilde {\mathcal{M}}_{x_1zx_2}-(2\pi)^4\frac{\lambda m^2}{6}\frac{1}{p_{x_1}p_{x_2}}|_{p_{x_1}^2=p_{x_2}^2=m^2}.
\end{split}
\end{equation}
We see that in the massless limit $\widetilde {\mathcal{N}}_{x_1 z , x_2}=\widetilde {\mathcal{M}}_{x_1zx_2}$ is recovered. It is not hard to see that exchanging $x_1\leftrightarrow x_2$ gives the same result, $\widetilde {\mathcal{N}}_{x_2 z , x_1}=\widetilde {\mathcal{N}}_{x_2 z , x_1}=\widetilde {\mathcal{M}}_{x_2zx_1}$. Therefore, let us compute the last term $\widetilde {\mathcal{N}}_{x_1 x_2 , z}$,
\begin{equation}
    \begin{split}
        \widetilde {\mathcal{N}}_{x_1 x_2 , z}&=\widetilde {\mathcal{M}}_{x_1x_2z}-\widetilde{\mathbf{\Gamma}}^w_{x_1x_2}\widetilde{\mathcal M}_{wz}\\
        &=\widetilde {\mathcal{M}}_{x_1x_2z}+(2\pi)^4\frac{\lambda p_z^2}{6}\left(\frac{1}{-p_zp_{x_1}}+\frac{1}{-p_{z}p_{x_2}}+\frac{1}{p_{x_2}p_{x_1}}\right)|_{p_{x_1}^2=p_{x_2}^2=m^2}\\
        &=\widetilde {\mathcal{M}}_{x_1x_2z}+(2\pi)^4\frac{\lambda (p_{x_1}+p_{x_2})^2}{6}\left(\frac{1}{(p_{x_1}+p_{x_2})p_{x_1}}+\frac{1}{(p_{x_1}+p_{x_2})p_{x_2}}+\frac{1}{p_{x_2}p_{x_1}}\right)|_{p_{x_1}^2=p_{x_2}^2=m^2}\\
        &=\widetilde {\mathcal{M}}_{x_1x_2z}+(2\pi)^4\frac{2\lambda (m^2+p_{x_1}p_{x_2})}{6}\left(\frac{1}{m^2+p_{x_1}p_{x_2}}+\frac{1}{m^2+p_{x_1}p_{x_2}}+\frac{1}{p_{x_2}p_{x_1}}\right)|_{p_{x_1}^2=p_{x_2}^2=m^2}\\
        &=\widetilde {\mathcal{M}}_{x_1x_2z}+(2\pi)^4\frac{\lambda}{3}\left(3+\frac{m^2}{p_{x_1}p_{x_2}}\right)|_{p_{x_1}^2=p_{x_2}^2=m^2}\\
        &=(2\pi)^4\frac{\lambda m^2}{3p_{x_1}p_{x_2}}|_{p_{x_1}^2=p_{x_2}^2=m^2},
    \end{split}
\end{equation}
where \eqref{eq: massive 3-pt} is used. Similarly, Eq. \eqref{eq: compat} is recovered in the massless limit, which justifies our earlier derivation using the metric compatibility condition.

Putting everything back into \eqref{eq: massive on-shell gamma} we get
\begin{equation}
    \begin{split}
        \widetilde \Gamma_{x_1 x_2}^y &= \frac{1}{2}\left(\widetilde {\mathcal{M}}^{-1}\right)^{y z}\left(\widetilde {\mathcal{N}}_{x_1 z , x_2}+\widetilde {\mathcal{N}}_{x_2 z , x_1}-\widetilde {\mathcal{N}}_{x_1 x_2 , z}\right)\\
        &=\frac{1}{2}\left(\widetilde {\mathcal{M}}^{-1}\right)^{y z}\left[\widetilde {\mathcal{M}}_{x_1zx_2}+\widetilde {\mathcal{M}}_{x_2zx_1}-(2\pi)^4\frac{\lambda}{3}\left(\frac{m^2}{p_{x_1}p_{x_2}}+\frac{m^2}{p_{x_1}p_{x_2}}\right)\right]\\
        &=\left(\widetilde {\mathcal{M}}^{-1}\right)^{y z}\left(\widetilde{\mathcal M}_{zx_1x_2}\right)-\lambda m^2 \left(\widetilde {\mathcal{M}}^{-1}\right)^{y z} \frac{(2\pi)^4}{3p_{x_1}p_{x_2}}\\
        &=G^y_{x_1x_2}-\lambda m^2 \left(\widetilde {\mathcal{M}}^{-1}\right)^{y z} \frac{(2\pi)^4}{3p_{x_1}p_{x_2}}.
    \end{split}
\end{equation}
In order to reproduce Eq.\eqref{eq:onshellconnect2}, we need to set $\lambda m^2=0$, and this validates our statement in Section \ref{sec: necessary conditions}.

\bibliographystyle{utphys}
\bibliography{ref}

@article{Vilkovisky:1984st,
    author = "Vilkovisky, G. A.",
    title = "{The Unique Effective Action in Quantum Field Theory}",
    doi = "10.1016/0550-3213(84)90228-1",
    journal = "Nucl. Phys. B",
    volume = "234",
    pages = "125--137",
    year = "1984"
}

@article{Finn:2020nvn,
    author = "Finn, Kieran and Karamitsos, Sotirios and Pilaftsis, Apostolos",
    title = "{Frame covariant formalism for fermionic theories}",
    eprint = "2006.05831",
    archivePrefix = "arXiv",
    primaryClass = "hep-th",
    reportNumber = "MAN/HEP/2020/004",
    doi = "10.1140/epjc/s10052-021-09360-w",
    journal = "Eur. Phys. J. C",
    volume = "81",
    number = "7",
    pages = "572",
    year = "2021"
}

@article{Gattus:2023gep,
    author = "Gattus, Viola and Pilaftsis, Apostolos",
    title = "{Minimal supergeometric quantum field theories}",
    eprint = "2307.01126",
    archivePrefix = "arXiv",
    primaryClass = "hep-th",
    doi = "10.1016/j.physletb.2023.138234",
    journal = "Phys. Lett. B",
    volume = "846",
    pages = "138234",
    year = "2023"
}

@inproceedings{DeWitt1985,
  author    = {Bryce S. DeWitt},
  title     = {The Effective Action},
  booktitle = {Architecture of Fundamental Interactions at Short Distances},
  series    = {Les Houches Summer School},
  year      = {1985},
  address   = {Les Houches, France},
  note      = {36 pp.},
}

@book{DeWitt1992Supermanifolds,
  author    = {Bryce S. DeWitt},
  title     = {Supermanifolds},
  edition   = {2nd, revised and expanded},
  series    = {Cambridge Monographs on Mathematical Physics},
  publisher = {Cambridge University Press},
  year      = {1992},
  address   = {Cambridge; New York},
  isbn      = {0-521-42377-5, 978-0-521-42377-9},
  pages     = {xviii + 407},
}

@article{Gattus:2024ird,
    author = "Gattus, Viola and Pilaftsis, Apostolos",
    title = "{Supergeometric quantum effective action}",
    eprint = "2406.13594",
    archivePrefix = "arXiv",
    primaryClass = "hep-th",
    doi = "10.1103/PhysRevD.110.105006",
    journal = "Phys. Rev. D",
    volume = "110",
    number = "10",
    pages = "105006",
    year = "2024"
}

@article{Alonso:2015fsp,
    author = "Alonso, Rodrigo and Jenkins, Elizabeth E. and Manohar, Aneesh V.",
    title = "{A Geometric Formulation of Higgs Effective Field Theory: Measuring the Curvature of Scalar Field Space}",
    eprint = "1511.00724",
    archivePrefix = "arXiv",
    primaryClass = "hep-ph",
    reportNumber = "CERN-PH-TH-2015-257",
    doi = "10.1016/j.physletb.2016.01.041",
    journal = "Phys. Lett. B",
    volume = "754",
    pages = "335--342",
    year = "2016"
}

@article{Alonso:2016oah,
    author = "Alonso, Rodrigo and Jenkins, Elizabeth E. and Manohar, Aneesh V.",
    title = "{Geometry of the Scalar Sector}",
    eprint = "1605.03602",
    archivePrefix = "arXiv",
    primaryClass = "hep-ph",
    reportNumber = "CERN-TH-2016-116",
    doi = "10.1007/JHEP08(2016)101",
    journal = "JHEP",
    volume = "08",
    pages = "101",
    year = "2016"
}

@article{Helset:2022tlf,
    author = "Helset, Andreas and Jenkins, Elizabeth E. and Manohar, Aneesh V.",
    title = "{Geometry in scattering amplitudes}",
    eprint = "2210.08000",
    archivePrefix = "arXiv",
    primaryClass = "hep-ph",
    reportNumber = "CALT-TH-2022-036",
    doi = "10.1103/PhysRevD.106.116018",
    journal = "Phys. Rev. D",
    volume = "106",
    number = "11",
    pages = "116018",
    year = "2022"
}

@article{Helset:2022pde,
    author = "Helset, Andreas and Jenkins, Elizabeth E. and Manohar, Aneesh V.",
    title = "{Renormalization of the Standard Model Effective Field Theory from geometry}",
    eprint = "2212.03253",
    archivePrefix = "arXiv",
    primaryClass = "hep-ph",
    reportNumber = "CALT-TH-2022-041",
    doi = "10.1007/JHEP02(2023)063",
    journal = "JHEP",
    volume = "02",
    pages = "063",
    year = "2023"
}

@article{Cohen:2021ucp,
    author = "Cohen, Timothy and Craig, Nathaniel and Lu, Xiaochuan and Sutherland, Dave",
    title = "{Unitarity violation and the geometry of Higgs EFTs}",
    eprint = "2108.03240",
    archivePrefix = "arXiv",
    primaryClass = "hep-ph",
    doi = "10.1007/JHEP12(2021)003",
    journal = "JHEP",
    volume = "12",
    pages = "003",
    year = "2021"
}

@article{Cohen:2020xca,
    author = "Cohen, Timothy and Craig, Nathaniel and Lu, Xiaochuan and Sutherland, Dave",
    title = "{Is SMEFT Enough?}",
    eprint = "2008.08597",
    archivePrefix = "arXiv",
    primaryClass = "hep-ph",
    doi = "10.1007/JHEP03(2021)237",
    journal = "JHEP",
    volume = "03",
    pages = "237",
    year = "2021"
}

@article{Derda:2024jvo,
    author = "Derda, Maria and Helset, Andreas and Parra-Martinez, Julio",
    title = "{Soft scalars in effective field theory}",
    eprint = "2403.12142",
    archivePrefix = "arXiv",
    primaryClass = "hep-th",
    reportNumber = "CALT-TH-2024-011, CERN-TH-2024-035",
    doi = "10.1007/JHEP06(2024)133",
    journal = "JHEP",
    volume = "06",
    pages = "133",
    year = "2024"
}

@article{Assi:2023zid,
    author = "Assi, Beno{\^\i}t and Helset, Andreas and Manohar, Aneesh V. and Pag{\`e}s, Julie and Shen, Chia-Hsien",
    title = "{Fermion geometry and the renormalization of the Standard Model Effective Field Theory}",
    eprint = "2307.03187",
    archivePrefix = "arXiv",
    primaryClass = "hep-ph",
    reportNumber = "CALT-TH-2023-024, FERMILAB-PUB-23-362-T",
    doi = "10.1007/JHEP11(2023)201",
    journal = "JHEP",
    volume = "11",
    pages = "201",
    year = "2023"
}

@article{Assi:2025fsm,
    author = "Assi, Beno{\^\i}t and Helset, Andreas and Pag{\`e}s, Julie and Shen, Chia-Hsien",
    title = "{Renormalizing Two-Fermion Operators in the SMEFT via Supergeometry}",
    eprint = "2504.18537",
    archivePrefix = "arXiv",
    primaryClass = "hep-ph",
    reportNumber = "CERN-TH-2025-084, FERMILAB-PUB-25-0280-V",
    month = "4",
    year = "2025"
}

@article{Alminawi:2025pwg,
    author = "Alminawi, Mohammad and Brivio, Ilaria and Davighi, Joe",
    title = "{Scalar Amplitudes from Fibre Bundle Geometry}",
    eprint = "2509.20482",
    archivePrefix = "arXiv",
    primaryClass = "hep-th",
    reportNumber = "COMETA-2025-38, CERN-TH-2025-187",
    month = "9",
    year = "2025"
}

@article{Alminawi:2023qtf,
    author = "Alminawi, Mohammad and Brivio, Ilaria and Davighi, Joe",
    title = "{Jet bundle geometry of scalar field theories}",
    eprint = "2308.00017",
    archivePrefix = "arXiv",
    primaryClass = "hep-ph",
    doi = "10.1088/1751-8121/ad72bb",
    journal = "J. Phys. A",
    volume = "57",
    number = "43",
    pages = "435401",
    year = "2024"
}

@article{Craig:2023wni,
    author = "Craig, Nathaniel and Lee, Yu-Tse and Lu, Xiaochuan and Sutherland, Dave",
    title = "{Effective field theories as Lagrange spaces}",
    eprint = "2305.09722",
    archivePrefix = "arXiv",
    primaryClass = "hep-th",
    doi = "10.1007/JHEP11(2023)069",
    journal = "JHEP",
    volume = "11",
    pages = "069",
    year = "2023"
}

@article{Cohen:2025prs,
    author = "Cohen, Timothy and Li, Xu-Xiang and Zhang, Zhengkang",
    title = "{Geometric Building Blocks of Effective Field Theory Amplitudes}",
    eprint = "2509.20449",
    archivePrefix = "arXiv",
    primaryClass = "hep-th",
    reportNumber = "CERN-TH-2025-188",
    month = "9",
    year = "2025"
}

@article{Cohen:2024bml,
    author = "Cohen, Timothy and Lu, Xiaochuan and Zhang, Zhengkang",
    title = "{What is the geometry of effective field theories?}",
    eprint = "2410.21378",
    archivePrefix = "arXiv",
    primaryClass = "hep-th",
    reportNumber = "CERN-TH-2024-176",
    doi = "10.1103/PhysRevD.111.085012",
    journal = "Phys. Rev. D",
    volume = "111",
    number = "8",
    pages = "085012",
    year = "2025"
}

@article{Cohen:2023ekv,
    author = "Cohen, Timothy and Lu, Xiaochuan and Sutherland, Dave",
    title = "{On amplitudes and field redefinitions}",
    eprint = "2312.06748",
    archivePrefix = "arXiv",
    primaryClass = "hep-th",
    reportNumber = "CERN-TH-2023-233",
    doi = "10.1007/JHEP06(2024)149",
    journal = "JHEP",
    volume = "06",
    pages = "149",
    year = "2024"
}

@article{Cohen:2022uuw,
    author = "Cohen, Timothy and Craig, Nathaniel and Lu, Xiaochuan and Sutherland, Dave",
    title = "{On-Shell Covariance of Quantum Field Theory Amplitudes}",
    eprint = "2202.06965",
    archivePrefix = "arXiv",
    primaryClass = "hep-th",
    doi = "10.1103/PhysRevLett.130.041603",
    journal = "Phys. Rev. Lett.",
    volume = "130",
    number = "4",
    pages = "041603",
    year = "2023"
}

@article{Craig:2025uoc,
    author = "Craig, Nathaniel and Lee, I-Kwan and Lee, Yu-Tse",
    title = "{Fermi Geometry of the Higgs Sector}",
    eprint = "2509.07101",
    archivePrefix = "arXiv",
    primaryClass = "hep-th",
    month = "9",
    year = "2025"
}

@article{Craig:2023hhp,
    author = "Craig, Nathaniel and Lee, Yu-Tse",
    title = "{Effective Field Theories on the Jet Bundle}",
    eprint = "2307.15742",
    archivePrefix = "arXiv",
    primaryClass = "hep-th",
    doi = "10.1103/PhysRevLett.132.061602",
    journal = "Phys. Rev. Lett.",
    volume = "132",
    number = "6",
    pages = "061602",
    year = "2024"
}

@article{Cheung:2022vnd,
    author = "Cheung, Clifford and Helset, Andreas and Parra-Martinez, Julio",
    title = "{Geometry-kinematics duality}",
    eprint = "2202.06972",
    archivePrefix = "arXiv",
    primaryClass = "hep-th",
    reportNumber = "CALT-TH 2022-006",
    doi = "10.1103/PhysRevD.106.045016",
    journal = "Phys. Rev. D",
    volume = "106",
    number = "4",
    pages = "045016",
    year = "2022"
}

@article{Berends:1987me,
    author = "Berends, Frits A. and Giele, W. T.",
    title = "{Recursive Calculations for Processes with n Gluons}",
    reportNumber = "Print-88-0100 (LEIDEN)",
    doi = "10.1016/0550-3213(88)90442-7",
    journal = "Nucl. Phys. B",
    volume = "306",
    pages = "759--808",
    year = "1988"
}

@article{Cohen:2025dex,
    author = "Cohen, Timothy and Fadakar, Ipak and Helset, Andreas and Nardi, Filippo",
    title = "{Geometry of soft scalars at one loop}",
    eprint = "2504.12371",
    archivePrefix = "arXiv",
    primaryClass = "hep-th",
    reportNumber = "CERN-TH-2025-081",
    doi = "10.1007/JHEP08(2025)140",
    journal = "JHEP",
    volume = "08",
    pages = "140",
    year = "2025"
}

@article{Aigner:2025xyt,
    author = "Aigner, Patrick and Bellafronte, Luigi and Gendy, Emanuele and Haslehner, Dominik and Weiler, Andreas",
    title = "{Renormalising the field-space geometry}",
    eprint = "2503.09785",
    archivePrefix = "arXiv",
    primaryClass = "hep-th",
    doi = "10.1007/JHEP07(2025)167",
    journal = "JHEP",
    volume = "07",
    pages = "167",
    year = "2025"
}

@article{Li:2024ciy,
    author = "Li, Xu-Xiang and Lu, Xiaochuan and Zhang, Zhengkang",
    title = "{The geometric universal one-loop effective action}",
    eprint = "2411.04173",
    archivePrefix = "arXiv",
    primaryClass = "hep-ph",
    doi = "10.1007/JHEP08(2025)102",
    journal = "JHEP",
    volume = "08",
    pages = "102",
    year = "2025"
}

@article{Lee:2024xqa,
    author = "Lee, Yu-Tse",
    title = "{Field space geometry and nonlinear supersymmetry}",
    eprint = "2410.21395",
    archivePrefix = "arXiv",
    primaryClass = "hep-th",
    doi = "10.1103/PhysRevD.111.105004",
    journal = "Phys. Rev. D",
    volume = "111",
    number = "10",
    pages = "105004",
    year = "2025"
}

@article{Criado:2018sdb,
    author = "Criado, J. C. and P{\'e}rez-Victoria, M.",
    title = "{Field redefinitions in effective theories at higher orders}",
    eprint = "1811.09413",
    archivePrefix = "arXiv",
    primaryClass = "hep-ph",
    doi = "10.1007/JHEP03(2019)038",
    journal = "JHEP",
    volume = "03",
    pages = "038",
    year = "2019"
}

@article{Cohen:2024fak,
    author = "Cohen, Timothy and Forslund, Matthew and Helset, Andreas",
    title = "{Field redefinitions can be nonlocal}",
    eprint = "2412.12247",
    archivePrefix = "arXiv",
    primaryClass = "hep-th",
    reportNumber = "CERN-TH-2024-216, YITP-SB-2024-35",
    doi = "10.1007/JHEP10(2025)019",
    journal = "JHEP",
    volume = "10",
    pages = "019",
    year = "2025"
}

@article{Arzt:1993gz,
    author = "Arzt, Christopher",
    title = "{Reduced effective Lagrangians}",
    eprint = "hep-ph/9304230",
    archivePrefix = "arXiv",
    reportNumber = "UM-TH-92-28",
    doi = "10.1016/0370-2693(94)01419-D",
    journal = "Phys. Lett. B",
    volume = "342",
    pages = "189--195",
    year = "1995"
}

\end{document}